\documentclass[twocolumn]{aastex631}
\usepackage{xspace}
\usepackage{amsmath}

\newcommand\scisub[2]{$#1_{\rm #2}$}

\newcommand{\hi}{H\,{\sc i}\xspace}
\newcommand{\himf}{H{\sc i}MF\xspace}

\newcommand{\mhi}{$M_\mathrm{HI}$\xspace} % mass of HI
\newcommand{\zhi}{$z_\mathrm{HI}$\xspace} % z of HI
\newcommand{\ohi}{$\Omega_{\rm HI}$\xspace} % omega of HI

\newcommand{\mstar}{$M_\ast$\xspace}
\newcommand{\phistar}{$\phi_\ast$\xspace}
\received{August 7, 2024}
\revised{December 12, 2024}
\accepted{December 12, 2024}

\begin{document}

\title{Looking At the Distant Universe with the MeerKAT Array: the \hi Mass Function in the Local Universe}

\author[0000-0003-4496-9553]{Amir Kazemi-Moridani}
\affiliation{Department of Physics and Astronomy, 
Rutgers, the State University of New Jersey, 
136 Frelinghuysen Road, Piscataway, NJ 08854, USA}

\author[0000-0002-7892-396X]{Andrew J. Baker}
\affiliation{Department of Physics and Astronomy, 
Rutgers, the State University of New Jersey, 
136 Frelinghuysen Road, Piscataway, NJ 08854, USA}
\affiliation{Department of Physics and Astronomy, University of the Western Cape, Robert Sobukwe Road, Bellville 7535, South Africa}

\author[0000-0001-9022-8081]{Marc Verheijen}
\affiliation{Kapteyn Astronomical Institute, University of Groningen, Landleven 12, 9747 AD, Groningen, The Netherlands}

\author[0000-0003-1530-8713]{Eric Gawiser}
\affiliation{Department of Physics and Astronomy, Rutgers, the State University of New Jersey, 136 Frelinghuysen Road, Piscataway, NJ 08854, USA}

\author[0000-0002-5777-0036]{Sarah-Louise Blyth}
\affiliation{Department of Astronomy, University of Cape Town, Private Bag X3, Rondebosch 7700, South Africa}

\author[0000-0002-1527-0762]{Danail Obreschkow}
\affiliation{International Centre for Radio Astronomy Research (ICRAR), The University of Western Australia, 35 Stirling Highway, Perth, WA 6009, Australia}

\author[0000-0002-3834-7937]{Laurent Chemin}
\affiliation{Universidad Andr\'es Bello, Facultad de Ciencias Exactas, Departamento de Ciencias F\'isicas -- Instituto de Astrof\'isica, Fernandez Concha 700, Las Condes, Santiago, Chile}

\author[0000-0002-2326-7432]{Jordan D. Collier}
\affiliation{Inter-University Institute for Data Intensive Astronomy (IDIA), Department of Astronomy, University of Cape Town, Private Bag X3, Rondebosch 7701, South Africa}
\affiliation{School of Science, Western Sydney University, Locked Bag 1797, Penrith NSW 2751, Australia}
\affiliation{CSIRO, Space and Astronomy, PO Box 1130, Bentley, WA 6102, Australia}

\author[0000-0002-4012-779X]{Kyle W. Cook}
\affiliation{Department of Physics and Astronomy, 102 Natural Science Building, University of Louisville, Louisville, KY 40292, USA}

\author[0000-0002-6149-0846]{Jacinta Delhaize}
\affiliation{Department of Astronomy, University of Cape Town, Private Bag X3, Rondebosch 7700, South Africa}

\author[0000-0001-9359-0713]{Ed Elson}
\affiliation{Department of Physics and Astronomy, University of the Western Cape, Robert Sobukwe Road, Bellville 7535, South Africa}

\author[0000-0003-3599-1521]{Bradley S. Frank}
\affiliation{The UK Astronomy Technology Centre, Royal Observatory Edinburgh, Blackford Hill, Edinburgh EH9 3HJ, UK}

\author[0000-0002-5067-8894]{Marcin Glowacki}
\affiliation{Institute for Astronomy, University of Edinburgh, Royal Observatory, Edinburgh EH9 3HJ, UK}
\affiliation{International Centre for Radio Astronomy Research (ICRAR), Curtin University, Bentley, WA 6102, Australia}
\affiliation{Inter-University Institute for Data Intensive Astronomy (IDIA), Department of Astronomy, University of Cape Town, Private Bag X3, Rondebosch 7701, South Africa}

\author[0000-0001-9662-9089]{Kelley M. Hess}
\affiliation{Department of Space, Earth and Environment, Chalmers University of 
Technology, Onsala Space Observatory, SE-43992 Onsala, Sweden}
\affiliation{ASTRON, the Netherlands Institute for Radio Astronomy, Postbus 2, 7990 AA, Dwingeloo, The Netherlands}

\author[0000-0002-4884-6756]{Benne W. Holwerda}
\affiliation{Department of Physics and Astronomy, 102 Natural Science Building, University of Louisville, Louisville, KY 40292, USA}

\author[0000-0002-8574-5495]{Zackary L. Hutchens}
\affiliation{Department of Physics \& Astronomy, University of North Carolina Asheville, 1 University Heights, Asheville, NC 28804, USA}

\author[0000-0001-7039-9078]{Matt J. Jarvis}
\affiliation{Astrophysics, University of Oxford, Denys Wilkinson Building, Keble Road, Oxford OX1 3RH, UK}
\affiliation{Department of Physics and Astronomy, University of the Western Cape, Robert Sobukwe Road, Bellville 7535, South Africa}

\author[0000-0002-1173-2579]{Melanie Kaasinen}
\affiliation{European Southern Observatory, Karl-Schwarzschild-Strasse 2, D-85748 Garching, Germany}

\author[0000-0001-9565-9622]{Sphesihle Makhathini}
\affiliation{Wits Centre for Astrophysics, School of Physics, University of the 
Witwatersrand, 1 Jan Smuts Avenue 2000, South Africa}

\author[0000-0001-6047-4521]{Abhisek Mohapatra}
\affiliation{Department of Astronomy, University of Cape Town, Private Bag X3, Rondebosch 7700, South Africa}

\author[0000-0002-9160-391X]{Hengxing Pan}
\affiliation{Astrophysics, University of Oxford, Denys Wilkinson Building, Keble Road, Oxford OX1 3RH, UK}

\author{Anja C. Schr\"oder}
\affiliation{Max-Planck-Institut f\"ur extraterrestrische Physik, Giessenbachstrasse 1, D-85748 Garching bei M\"unchen, Germany}

\author{Leyya Stockenstroom}
\affiliation{Department of Astronomy, University of Cape Town, Private Bag X3, Rondebosch 7700, South Africa}

\author[0000-0002-6748-0577]{Mattia Vaccari}
\affiliation{Inter-University Institute for Data Intensive Astronomy, Department of Astronomy, University of Cape Town, Private Bag X3, 7701 Rondebosch, Cape Town, South Africa}
\affiliation{Inter-University Institute for Data Intensive Astronomy, Department of Physics and Astronomy, University of the Western Cape, 7535 Bellville, Cape Town, South Africa}
\affiliation{INAF -- Istituto di Radioastronomia, via Gobetti 101, 40129 Bologna, Italy}

\author[0000-0002-5300-2486]{Tobias Westmeier}
\affiliation{International Centre for Radio Astronomy Research (ICRAR), The 
University of Western Australia, 35 Stirling Highway, Perth, WA 6009, Australia}

\author[0000-0002-5077-881X]{John F. Wu}
\affiliation{Space Telescope Science Institute, 3700 San Martin Drive, Baltimore, MD 21218-2410, USA}
\affiliation{Center for Astrophysical Sciences, Johns Hopkins University, 3400 North Charles Street, Baltimore, MD 21218-2608, USA}

\author[0000-0003-0101-1804]{Martin Zwaan}
\affiliation{European Southern Observatory, Karl-Schwarzschild-Strasse 2, D-85748 Garching, Germany}

%\author{et al.}



\begin{abstract}

We present measurements of the neutral atomic hydrogen (\hi) mass function (\himf) and cosmic \hi density (\ohi) at $0\leq z \leq 0.088$ from the Looking at the Distant Universe with MeerKAT Array (LADUMA) survey. Using LADUMA Data Release 1 (DR1), we analyze the \himf via a new ``recovery matrix'' (RM) method that we benchmark against a more traditional Modified Maximum Likelihood (MML) method. Our analysis, which implements a forward modeling approach, corrects for survey incompleteness and uses extensive synthetic source injections to ensure robust estimates of the \himf parameters and their associated uncertainties. This new method tracks the recovery of sources in mass bins different from those in which they were injected and incorporates a Poisson likelihood in the forward modeling process, allowing it to correctly handle uncertainties in bins with few or no detections. The application of our analysis to a high-purity subsample of the LADUMA DR1 spectral line catalog in turn mitigates any possible biases that could result from the inconsistent treatment of synthetic and real sources. For the surveyed redshift range, the recovered Schechter function normalization, low-mass slope, and ``knee'' mass are $\phi_\ast = 3.56_{-1.92}^{+0.97} \times 10^{-3}$ Mpc$^{-3}$ dex$^{-1}$, $\alpha = -1.18_{-0.19}^{+0.08}$, and $\log(M_\ast/M_\odot) = 10.01_{-0.12}^{+0.31}$, respectively, which together imply a comoving cosmic \hi density of \ohi$=3.09_{-0.47}^{+0.65}\times 10^{-4}$. Our results show consistency between RM and MML methods and with previous low-redshift studies, giving confidence that the cosmic volume probed by LADUMA, even at low redshifts, is not an outlier in terms of its \hi content.
\end{abstract}

\keywords{Galaxies (573); Galaxy masses (607); \hi line emission (690)}


\section{Introduction} \label{sec:intro}

Hydrogen, the most abundant element in the Universe, exists in a variety of gaseous phases spanning broad ranges of temperature and density. Neutral atomic hydrogen (\hi) acts as a crucial intermediate stage in the evolution of gas in galaxies, bridging the gap between ionized hydrogen flowing in from the intergalactic medium and molecular hydrogen, which serves as the primary fuel for star formation \citep{haynes84,mccg23}. \hi masses in galaxies can change due to the processes of accretion, consumption/conversion, and expulsion. Consequently, tracking the evolution of \hi over cosmic time is vital for understanding galaxy evolution. The number density of galaxies as a function of \hi mass (i.e., the neutral hydrogen mass function or \himf) and its integral, the contribution of galaxies to the cosmic \hi density (\ohi), are two key metrics for describing the distribution and abundance of neutral gas across cosmic time. Measurements of the \himf and \ohi enable comparisons with semi-analytic and numerical models of galaxy evolution \citep{Popping14, Kim15, Dave20}.

Beyond cosmic averages, understanding \hi properties in different settings can illuminate the dependence of galaxy evolution processes on environmental factors. \hi is particularly sensitive to galaxy interactions as it is affected by hydrodynamic pressure and has a more broadly extended distribution within galaxies compared to stars and other gas phases \citep[e.g.,][]{mihos01}. The \hi contents and distributions of galaxies reflect histories of interaction and the impacts of multiple evolutionary mechanisms, with \hi surveys providing vital data for validating simulations and understanding the environmental variations of galaxy evolution \citep[e.g.,][]{chung09,holwerda11,jones18,reynolds22}.

The \hi 21\,cm line, resulting from a hyperfine transition in the ground state of hydrogen atom, serves as an essential tool for investigating neutral hydrogen on Galactic and extragalactic scales. Over the past few decades, observations of the 21\,cm line have revolutionized our understanding of \hi distributions within the local universe and beyond. Numerous \hi emission line surveys, conducted with single-dish radio telescopes \citep{zwaan03, zwaan05, martin10, jones18}, 
have played a pivotal role in assessing the global \hi properties of nearby galaxies. The two largest untargeted \hi surveys, the \hi Parkes All-Sky Survey \citep[HIPASS;][]{Meyer04} and the Arecibo Legacy Fast ALFA \citep[ALFALFA;][]{giovanelli05} survey, have measured the \himf and \ohi in the nearby universe. These surveys and the subsequent interferometric analysis of \citep{pono23} have been instrumental in establishing our knowledge of the \himf and its properties, including the contributions of various galaxy populations to the \himf and its dependence on environment \citep{moorman14, said19, jones20}. 

Previous \hi surveys have predominantly focused on sources in the local ($z <0.1$) Universe due to the faintness (i.e., low Einstein A coefficient) of the 21\,cm line and the limited capabilities of the telescopes that observe it. Despite significant investments of observing time, so far only a small number of galaxies have been detected 
in \hi beyond the local Universe \citep{fern13, fern16, highz15, hess19, gogate20}, with the most distant individual detection at $z \approx 0.42$ \citep{xi24}. Only two surveys have managed to assess the \himf beyond the $z <0.1$ Universe; the Arecibo Ultra-Deep Survey \citep[AUDS; ][]{hoppmann15, xi21} covers $0 <z <0.16$, while the Blind Ultra-Deep \hi Environmental Survey \citep[BUDHIES;][]{jaffe13, gogate20} has for the first time used direct \hi detections to construct the \himf and calculate \ohi at $z \sim 0.2$ \citep{avantiThesis}, in two volumes centered on galaxy clusters. 

Although compiling a large sample of direct \hi detections at higher redshifts for statistical investigations remains an elusive goal, several ambitious surveys are aiming to address this deficiency, using new facilities such as the Australian Square Kilometre Array Pathfinder \citep[ASKAP;][]{hotan21}, the Five-hundred-meter Aperture Spherical radio Telescope \citep[FAST;][]{nan08}, and the MeerKAT array \citep{jonas09}, and the phased array feed (PAF) upgrade of the Westerbork Synthesis Radio Telescope known as the APERture Tile In Focus \citep[APERTIF;][]{adams22}. This next generation of untargeted \hi surveys is set to enhance our understanding of how the \himf and \ohi evolve over cosmic time. Surveys such as the MeerKAT International GigaHertz Tiered Extragalactic Exploration \citep[MIGHTEE;][]{jarvis16,maddox21} and the Deep Investigation of Neutral Gas Origins \citep[DINGO;][]{rhee23} are designed to probe \hi in galaxies across a broad range of redshifts and environments. Building on the precedent of the deep but narrow COSMOS \hi Legacy Extragalactic Survey \citep[CHILES;][]{fern16,hess19} with the Very Large Array (VLA), 
the Looking At the Distant Universe with the MeerKAT Array \citep[LADUMA;][]{blyth18} survey now aims to explore \hi in emission up to an unprecedented $z \sim 1.4$, utilizing both direct and stacked detections, and is designed to detect thousands of galaxies at its targeted depth.

Previous analyses have inferred notable discrepancies in the shape of the \himf in the local universe \citep[e.g., between ALFALFA's spring and fall volumes;][]{jones18}, but it is unclear whether these variations 
reflect fundamental differences in galaxy evolution in different environments or are expected consequences
of cosmic variance. As we extend our observational reach with the current generation of \hi surveys, the catalogs curated at higher redshifts are expected to contain few detections compared to surveys of the local Universe, further increasing the difficulties of drawing robust conclusions and motivating the development of improved analytical methods. To fully leverage the potential of these upcoming samples and to address some of the observed discrepancies at lower redshifts, it is imperative to refine and develop new methods for deriving the \himf. 

The present paper addresses this critical need by proposing a novel approach to improve the analysis and interpretation of \himf estimates across different redshifts. We demonstrate its application by presenting the first determination of the \himf using observations from LADUMA. Using LADUMA's Data Release 1 (DR1) dataset, we derive the \himf over roughly the last billion years ($0 < z < 0.088$) and calculate an associated \ohi for this redshift range. The paper is organized as follows. In Section \ref{sec:data}, we describe the LADUMA survey, our processing of the DR1 data, and our definition of a high-purity \hi sample. In Section \ref{sec:analysis}, we describe how we measure the \himf, which is parameterized in terms of a \citet{schechter76} function; the results of our analysis are presented in Section \ref{sec:res}. We discuss the novel aspects of our method and implications for its application at higher redshift in Section \ref{sec:disc}. A summary and conclusions are presented in Section \ref{sec:sum}. Throughout this paper, we assume a flat cold dark matter ($\Lambda$CDM) cosmology with $H_0$ = 70 km~s$^{-1}$~Mpc$^{-1}$, $\Omega_m$ = 0.3, and $\Omega_\Lambda$ = 0.7. All reported values from the literature have been rescaled as necessary for consistency with this cosmology.

\section{Observations} \label{sec:data}
\subsection{Data Processing} 
LADUMA is a deep untargeted 21\,cm survey of a single pointing using the MeerKAT array, with an area that expands from $\sim 0.8~{\rm deg}^2$ at \zhi $= 0$ to $\sim 5~{\rm deg}^2$ at \zhi $= 1.4$ \citep{blyth18}. The LADUMA pointing encompasses the Chandra Deep Field South (CDFS) and is centered at 03:32:30 $-$28:07:57 (J2000). The LADUMA data were processed on the ilifu facility operated by the Inter-University Institute for Data Intensive Astronomy (IDIA), using the {\tt processMeerKAT} pipeline\footnote{\url{https://idia-pipelines.github.io/docs/processMeerKAT}} \citep{Collier21} combined with custom scripts.
The data used in this paper were collected in 19 night-time tracks with MeerKAT's L-band (0.88--1.67\,GHz) receivers, whose average duration was 9 hours. Data were obtained at the native 32k channel resolution ($26.123$~kHz) of the MeerKAT correlator and processed at 8k resolution in three independent spectral windows (SPWs) of 880-933 MHz, 960-1161 MHz, and 1304-1420 MHz, in order to avoid regions of strong radio frequency interference. This paper focuses on the data in the highest-frequency (low-$z$) SPW, which corresponds to 0 $<$ \zhi $<$ 0.088. 

\begin{figure*}
\centering
\includegraphics[width=0.99\textwidth]{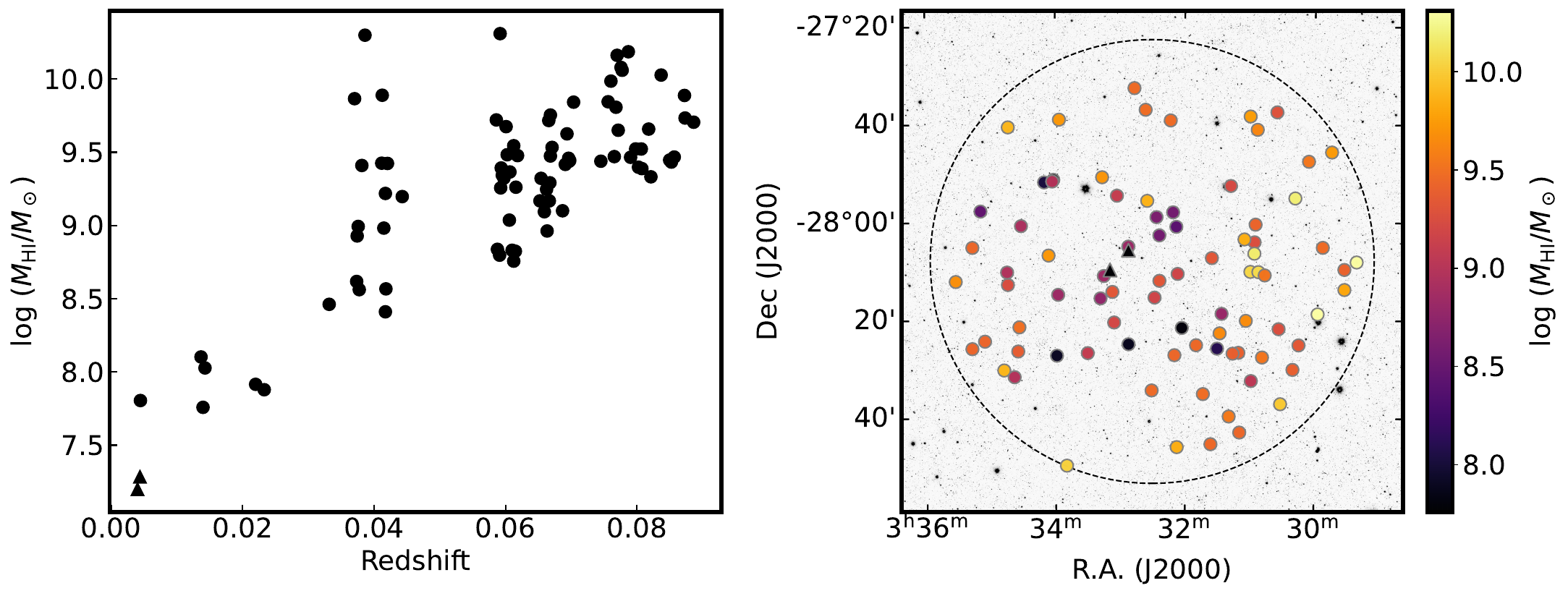}
\caption{LADUMA \hi sources detected by SoFiA with higher than $99\%$ reliability inside the FWQM (two sources with $M_{\rm H\,I} < 10^{7.5}\,M_\odot$, which are not used in our analysis, are plotted as triangles). Left: \hi mass as a function of redshift. Right: High-purity subsample of \hi sources in the LADUMA field plotted overlaid on the Spitzer/IRAC channel 1 mosaic for the extended CDFS \citep{Euclid22}. The dashed circle represents the FWQM of the MeerKAT primary beam at 1304 MHz (the lower frequency edge of the spectral window analyzed here).
\label{fig:srcs}}
\end{figure*}

The full width at half maximum (FWHM) of the MeerKAT primary beam\footnote{The primary beam response is calculated using the \url{https://github.com/ska-sa/katbeam} package.} expands from 60\arcmin.5 at 1420 MHz to 65\arcmin \xspace at 1304 MHz. The outstanding sensitivity of MeerKAT allows for the detection of \hi emitters well past the FWHM of the primary beam. As a result, in this work we analyze a volume extending to the full width at quarter maximum (FWQM) of the primary beam, which increases from 83\arcmin \xspace to 90\arcmin \xspace over the frequency range of the SPW. A detailed account of the data reduction will be presented in a forthcoming paper (Kazemi-Moridani et al., in preparation), but we summarize key steps here.

Each track is individually calibrated and imaged, with robust weighting adjusted per track to minimize sidelobes. We combine the tracks after subtracting a continuum sky model from each in the $uv$ plane. At its central frequency, the combined data cube has synthesized beam dimensions of $8.^{\prime\prime}0 \times 7.^{\prime\prime}5$ with ${\rm P.A.} = -34^\circ$ and an RMS noise of $\sim$33~$\mu$Jy~beam$^{-1}$ per 104.52\,kHz channel. Model subtraction leaves low-level continuum residuals, especially in the vicinity of bright continuum sources, that are not visible in a single-track image. However, as the noise integrates down with the combination of multiple tracks, the remaining continuum residuals emerge above the lower noise. We have developed a spline\footnote{We use splines rather than polynomials because splines better capture the smooth, ripple-like behavior in the pixel spectra.}-fitting algorithm to remove these residuals in our deepest data cube, which we will refer to as pixel-based continuum subtraction from this point on. This latter stage of continuum subtraction affects our measured \hi fluxes, as it models the underlying continuum emission at locations where the line sources reside. Overestimation (underestimation) of the continuum level can result in underestimation (overestimation) of the line flux, as can also occur for alternative approaches to continuum subtraction \citep[e.g.,][]{Meyer04, haynes11}. For the LADUMA DR1 cubes, this effect is minimal for sources with small velocity widths, as the spline-fitting algorithm only needs to model the continuum level over a few channels. For sources with larger velocity widths, the specific choices made in implementing the pixel-based continuum subtraction algorithm have the potential to systematically affect the total measured line fluxes, albeit at a modest (typically $< 5-10\%$) level. 

\subsection{Source catalog} \label{ssec:sf}

Given the need for an automated source-finder for injection/recovery tests (see Section \ref{ssec:injec} below), we make use of the Source Finding Application \citep[SoFiA;][]{westm21} package, which offers excellent efficiency, flexibility, and reliability. Our source-finding approach is refined by maximizing the fraction of synthetic sources recovered with high fidelity in the entirety of the synthetic source population. To maintain consistency between finding real sources and finding synthetic sources, we apply the same approach to the real data and use a high-purity subsample of all the line sources detected in the low-$z$ SPW. We note that this high-purity subsample is considerably smaller than the full LADUMA DR1 source catalog. That catalog, which will be described in detail elsewhere (Kazemi-Moridani et al., in preparation), includes both a core sample (for which SoFiA parameters were iteratively adjusted to match the results of unguided source-finding with visual and matched-filtering methods) and a supplemental sample (containing the results of visual source-finding guided by prior knowledge of optical redshifts, as well as a more extensive exploration of SoFiA parameter space).  Sources in the supplemental sample generally have lower S/N and therefore do not figure in the high-purity subsample used in this paper.  The SoFiA parameters we use in this analysis match those used to develop the core DR1 sample and are listed in Table \ref{tab:sofia} (if different from package defaults).  For these parameters, in the low-$z$ SPW data cube, SoFiA finds $\sim 190$ candidate sources, of which we judge $\sim 140$ to be real detections based on visual inspection and cross-matching with optical catalogs.  
On the basis of previous predictive work by \citet{roberts21}, and given the low redshifts in question, we do not expect any OH megamasers (OHMs) to be present in this list.

Out of our parent sample of $\sim 140$ sources, 89 are detected with high enough reliability ($> 99\%$; see \S \ref{ssec:injec} below) to be included in the high-purity sample. By limiting our analysis to the FWQM of the primary beam, we reduce the number of included detections from 89 to 84, ensuring that only high-SNR massive sources remain. Furthermore, in order to avoid biasing the \himf by relying on the very small volume associated with the two lowest-mass sources (i.e., sources in the lowest-mass bin; see Section \ref{ssec:comp} for further details), we exclude them from our analysis as well, leaving 82 sources in the final sample used to measure the \himf. Figure \ref{fig:srcs} shows the distribution of these 82 sources (see Section \ref{ssec:injec}) as a function of redshift and on the plane of the sky.
Given the rarity of rich clusters \citep[e.g.,][]{Abell58} and the deficiency of \hi emission in the densest environments \citep[e.g.,][]{Giovanelli85}, we expect that our \hi-selected sources are primarily located in field or intermediate density environments. Considering LADUMA's angular resolution, we also expect that spectral line confusion will not significantly affect the recovered source population \citep[e.g.,][]{Jones15}.

\begin{table*}[h]
\centering
\caption{SoFiA parameters used for this paper's injection/recovery analysis. Parameters are only listed when adopted values differ from defaults.}
\label{tab:sofia}
% \tabletypesize{8pt}
\begin{tabular}{ccc}
\hline
Parameter &  Default & Used here \\
\hline
window$\_$x & 25 & 63 \\
window$\_$y & 25 & 63 \\
window$\_$z & 15 & 31 \\
scalenoise.interpolate & false & true \\
scfind.kernelsXY & 0, 3, 6 & 0, 2, 5 \\
scfind.kernelsZ & 0, 3, 7, 15 & 0, 3, 5, 9 \\
threshold & 5 & 4 \\
linker.minsizeZ & 5 & 3 \\
linker.radiusXY & 1 & 2 \\
linker.radiusZ & 1 & 2 \\
reliability.threshold & 0.9 & 0.7 \\
reliability.minSNR & 3 & 5 \\
\hline
\end{tabular}
\end{table*}

\section{Analysis} \label{sec:analysis}

\subsection{Recovery matrix method}

\subsubsection{Injecting and recovering sources} \label{ssec:injec}
The \himf is defined as the number density of galaxies as a function of \hi mass per unit comoving volume, which is usually represented as
\begin{equation}
\phi(M_{\rm HI})=\frac{dN_{\rm gal}}{dV\,d\log_{10}(M_{\rm HI}) }
\end{equation}
where $dN_{\rm gal}$ is the number of galaxies with \hi masses falling in a logarithmic mass bin centered on \mhi and lying in the comoving volume $dV$. Determining the intrinsic \himf from observed number counts is a complex problem, especially when a sample is not volume-limited.
To measure the \himf, it is essential to correct for the survey sensitivity, which requires estimating the completeness of the observed sample. Completeness is typically calculated for each \hi mass bin as an estimate of the fraction of all galaxies within the associated mass range that have actually been detected. The main factors affecting completeness in our data are distance, non-uniform sensitivity across our field of view due to primary beam attenuation, random orientations of individual \hi emitters on the sky (affecting the observed linewidths of those sources), uncertainties in flux measurements, continuum subtraction effects, and source-finding accuracy. The combined effects of these factors bias our observed \hi galaxy sample towards nearby, gas-rich, and low-inclination sources near the center of the LADUMA field. To assess the completeness of our survey, we adopt an empirical approach that involves inserting synthetic sources into our processed data cube. We evaluate their recovery rate by requiring that they be detected via the same process as the real detections in our survey. As described in Section \ref{ssec:sf} and given the number of artificial source injections required to accurately sample the multi-dimensional parameter space of real \hi sources, employing this approach requires a highly automated source-finder that can detect galaxies with high reliability. 

All injection/recovery methods are based on prior knowledge of the underlying distributions of source parameters, such as mass, inclination, size, and velocity width. To create a catalog of synthetic sources that accurately samples the multi-dimensional parameter space spanned by \hi sources, we follow an approach based the one detailed in \citet{gogate20}. The synthetic \hi sources are simulated using the GIPSY \citep[Groningen Image Processing System;][]{hulst92} package's {\tt galmod} task, which uses a tilted ring model \citep{rogstad74} to create the 3D velocity field of a given \hi source. For each galaxy, {\tt galmod} requires the radial \hi surface density distribution (in cm$^{-2}$) and the rotational velocity distribution (in km~sec$^{-1}$) as a function of the radius (in arcseconds) of the galaxy, as well as the velocity dispersion (in km~sec$^{-1}$), the inclination angle, and the position angle (in degrees). We use the radial surface density distribution profiles as described in \citet{Sersic68} and \citet{Martinsson16} and the rotational velocity profiles described by \citet{Persic96} and \citet{Courteau97}, as applicable to different mass ranges, to generate synthetic \hi sources across the wide mass range of our study. We make use of the known local \hi size-mass scaling relation \citep{wang16} when generating the parameters required by {\tt galmod} to guarantee that the sources follow this relation. The equation for the amplitude of the rotational velocity profiles \citep[from][]{Persic96} likewise yields results that are consistent with the Tully-Fisher relation \citep{TF77}. Our final catalog comprises over 150k synthetic sources, with randomly chosen inclinations and position angles, covering \hi masses in the range 7 $\leq$ log (\mhi/$M_\odot$) $<$ 10.75, such that the vast majority of sources have masses in the lower half of the range as required to ensure reliable recovery statistics for the entire mass range (see below). To ensure minimal alterations to the noise properties of the data cube, which can influence the source-finding process, the number of sources in any single injection/recovery trial is limited to a maximum of 500 for sources near the low-mass limit and a maximum of 100 for sources near the high-mass limit.

When SoFiA is run on a given cube that includes both real and synthetic sources, it delivers a list of all spectral line detections and --- for each detection --- a three-dimensional ``cubelet'' that includes generous spatial and spectral buffers around the pixels in which emission is seen.\footnote{The SoFiA parameters used for this stage of our analysis differ from those used to produce the original source catalog only in allowing the detection of sources with very large angular sizes, as is appropriate for synthetic sources with large \hi masses and low redshifts.} We calculate the total \hi mass of each detection as
\begin{equation}
\label{eq:mass}
\left(\frac{M_{\rm HI}}{M_\odot}\right) = 49.7 \left(\frac{D_{L}}{{\rm Mpc}}\right)^{2}\left(\frac{S}{{\rm Jy~Hz}}\right),
\end{equation}
where $D_L$ is the cosmological luminosity distance to the source and $S$ is the integrated \hi flux density \citep{meyer17}. The integrated \hi flux density is calculated using zeroth moment maps that are created by first smoothing individual source cubelets to a circular beam of $20^{\prime\prime} \times 20^{\prime\prime}$. Each smoothed cubelet is clipped at a 3$\sigma$ threshold ($\sigma$ is estimated by measuring the RMS noise over an emission-free region) to create a mask that encompasses diffuse low-column-density emission. We remove isolated regions corresponding to noise peaks from the mask by discarding all regions whose areas are smaller than that of the smoothed beam. We then apply the resulting mask to the original-resolution cubelet to generate the zeroth moment map and calculate the integrated \hi flux density.

False positive detections can become a major source of error in recovery rate calculations. Given our perfect knowledge of where the synthetic sources are, we can easily identify any false positives that SoFiA ``recovers''; however, it is not possible to similarly distinguish false from real sources in our full observed sample. In order to eliminate the complications that would be caused by false positive detections in our catalog of real sources, we need to work only with a high-purity subset of SoFiA detections in both the observed and synthetic data. Recovery tests on the synthetic sources have shown that using SoFiA's reliability threshold of 99\% creates a high-purity sub-sample (purity $>$ 99\%).\footnote{Because no analogs of multiwavelength counterparts exist for synthetic sources, in contrast to real sources, we cannot incorporate the use of multiwavelength catalogs into the source-finding process for real sources. Without using multiwavelength data, it becomes much more challenging to determine how many source candidates identified by SoFiA (e.g., above a lower reliability threshold of 95\%) are false detections. Therefore, we need to establish a reliability threshold using injection recovery tests alone that delivers high-purity samples (e.g., the high-purity sample from the 99\% reliability threshold), which do not require cross-validation against multiwavelength data.} All the real sources in the selected sub-sample are detected in every injection/recovery trial run, which confirms that the injected population in each trial does not alter source-finding accuracy. 
Based on optical redshift measurements and WISE-based color diagnostics \citep{roberts21}, our catalog does not include any detections of OH megamasers.
We note that our high-purity sample of 84 \hi detections is considerably smaller than the full DR1 catalog of \hi detections (Kazemi-Moridani et al., in preparation); this circumstance reflects our deliberate choice to prioritize the accurate determination of sample completeness over a large sample size, which are inescapably in tension with each other.

\begin{figure}
\centering
\includegraphics[width=0.48\textwidth]{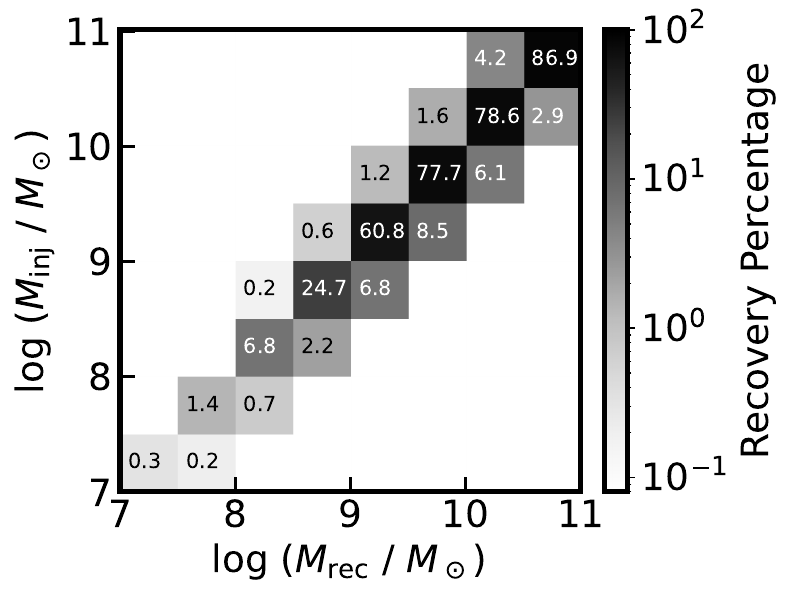}
\caption{The recovery fraction matrix plotted on the \scisub{M}{inj} vs. \scisub{M}{rec} grid. The diagonal elements express recovery fraction as a percentage and increase monotonically as a function of injected mass \scisub{M}{inj} and recovered mass \scisub{M}{rec}. For the lower-mass bins, the off-diagonal elements are large relative to the on-diagonal elements. If a source is not recovered in the injected mass bin, it is more likely to be recovered with a higher mass than with a lower mass.
\label{fig:rec_mat}}
\end{figure}

Each injection/recovery trial includes the following steps:
\begin{itemize}
    \item creating synthetic sources in the desired mass range (\mhi is randomly sampled from a log-uniform distribution and inclinations are sampled from a $p(i)di = \sin(i) di$ distribution) with {\tt galmod}\footnote{The sources are spectrally smoothed in {\tt galmod} with a Gaussian kernel whose FWHM is twice the channel separation.};
    \item convolving the {\tt galmod} output for each source with the point spread function (PSF) of our data as appropriate for the assumed source redshift (redshifts are sampled from a $p(z) \propto dV(z)/dz$ distribution, where $V(z)$ represents the volume of the survey as a function of redshift\footnote{Accurately capturing the effects of large-scale structure (LSS) at the point of source injection would (ideally) entail simultaneously solving for both the underlying LSS and the \himf\ using our observed mass and spatial distributions. Unfortunately, our sample size is insufficient to support such a joint analysis. Instead, we have assessed source recovery statistics in the presence of LSS analogous to that observed in our real source catalog. These tests were focused on sources with ${\rm log}\,(M_{\rm HI}/M_\odot) \geq 9$, which represent 75\% of our observed sources and sample volumes where a reasonable estimate of LSS can be derived from \hi\ data. Our tests confirm that the presence of LSS in {\it our} data along the line of sight does not significantly affect source recovery statistics as a function of \hi\ mass, although other surveys might be affected differently.});
    
    \item injecting sources in the pre-pixel-based-continuum-subtraction data cube at randomly selected locations with a uniform cosmological volume density distribution (while avoiding overlap with existing sources in the full LADUMA catalog\footnote{The spatial position of a given synthetic source is determined in ($r$, $\theta$) coordinates with respect to the pointing center. Radius $r$ is sampled according to $p(r)dr = rdr$, and $\theta$ is sampled from a uniform distribution between 0 and 2$\pi$. Each ($r_i$, $\theta_i$, $z$) tuple is cross-checked with existing source positions to avoid overlap.});
    \item running the second-stage (pixel-based) continuum subtraction algorithm on the data cube, including both real and synthetic sources;
    \item running SoFiA on the resulting continuum-subtracted line cube to find spectral line sources;
    \item cleaning all recovered sources using a custom \cite{hogbom74} CLEAN  routine applied to the SoFiA masks; and
    \item cross-matching the (RA, Dec, $z$) tuples of the recovered source catalog with the (RA, Dec, $z$) tuples of the input source catalog.\footnote{99\% of the injected sources are recovered within a sphere with a 3.5 spaxel radius centered on the injected position.}
\end{itemize}

The outputs of our 500 trials are then combined to create a reference recovery catalog in which the fate of each injected source and its properties are recorded. We split the total mass range (7 $\leq$ log (\mhi/$M_\odot$) $\leq$ 11) into eight bins in order to investigate the recovery details for each bin. A traditional approach here would calculate the ratio of recovered to injected sources in each bin and correct the number of real detections by the recovery factor to determine the underlying distribution of galaxies. However, our simulations show that even with a recovery fraction of close to 100\%, a considerable number of galaxies injected in a given mass bin are recovered in a different bin, with a shift into a higher-mass bin much more likely than the alternative. Given that there are significantly more low-mass objects than higher-mass objects in the real universe, the recovery of sources in higher-mass bins can introduce a bias in the inferred \himf that needs to be corrected. 

\subsubsection{Defining a recovery matrix}

In order to keep track of the migration of recovered sources across mass bins, we create a correction matrix --- a 2D array of recovery factors (instead of a correction vector, i.e., a 1D array of recovery factors) --- in which the fates of sources injected in each mass bin are recorded as a function of their recovered mass. This ``recovery matrix'' can be thought of as a function over a \scisub{M}{inj(ected)} vs. \scisub{M}{rec(overed)} domain, where the function at a given (\scisub{M}{inj}$, $\scisub{M}{rec}) point represents the fraction of sources with an injected mass in the \scisub{M}{inj} bin that have been recovered with a mass in the \scisub{M}{rec} bin. In an ideal case, the recovery fraction matrix would be the identity matrix, signifying that each source has been recovered in the same mass bin in which it was injected. Our tests show that the recovery fraction matrix for our data cube is a band-diagonal matrix with significant nonzero values on the main diagonal and the adjacent diagonals, such that sources are most likely to be recovered in their original injection bin and much more likely to be recovered in the next-higher mass bin than in the next-lower mass bin (Figure \ref{fig:rec_mat}). The greater likelihood of recovering sources in the adjacent higher-mass bin results from the combined effect of several factors, notably the effects of the source-finding process and image-plane continuum subtraction. Source finding relies heavily on threshold cuts, such that a source near the detection limit might be selected as a detection if it coincides with a positive noise region, but ignored if it coincides with a negative noise region.\footnote{We note that even high-mass sources can lie close to the detection threshold if they lie at large redshifts and/or large offsets from the pointing center. This positive bias is different from the 2--3\% positive bias in SoFiA flux measurements that is noted by \citet{westm21}; the latter is not relevant to our analysis here because we do not use SoFiA to measure fluxes (see above).} Estimation of the underlying continuum during pixel-based continuum subtraction in turn could be biased toward underestimating the continuum level for some sources depending on their parameters (such as inclination), leading to under-subtraction of continuum and overestimated line fluxes for those sources. The combination of these factors, along with the effects of other steps in our data processing, leads to an asymmetric bias towards recovery at higher masses.

\begin{figure}
\centering
\includegraphics[width=0.5\textwidth]{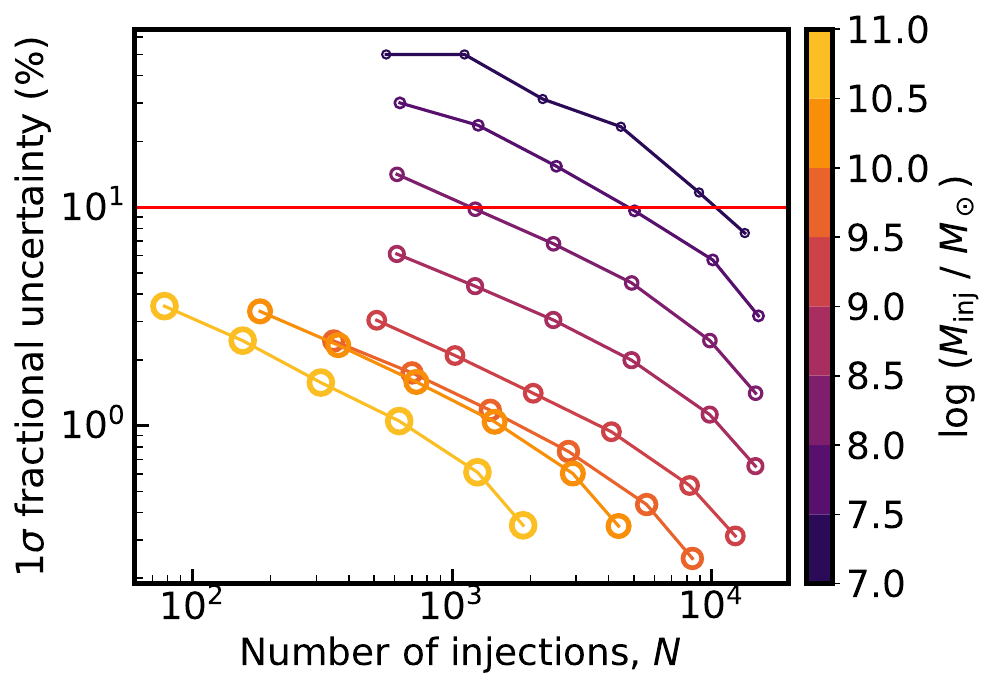}
\caption{1$\sigma$ fractional uncertainty (as a percent) on the diagonal elements of the recovery fraction matrix shown in Figure \ref{fig:rec_mat} as a function of the number of injections $N$. The fractional uncertainty is determined by calculating the recovery matrix elements for 4096 randomly chosen samples of $N$ sources in a given \scisub{M}{inj} bin from a larger pool of injected sources in that bin. The final 1$\sigma$ fractional uncertainties on all the diagonal (shown here) and below-diagonal matrix elements are less than 10\%.
\label{fig:diag_el}}
\end{figure}

In order to make sure that we have an accurate estimate of each matrix element, we have injected enough sources to ensure that the fractional uncertainties on the diagonal and below-diagonal matrix elements are less than 10\%. Figure \ref{fig:diag_el} shows the percentage uncertainties on the diagonal matrix elements as a function of the number of injected sources. To calculate the uncertainties on the matrix elements for the recovery of synthetic \hi sources, we employ a method inspired by the bootstrapping technique. For each mass bin in the analysis, we start with a large pool of synthetic sources --- specifically, at least 15\% more sources than the maximum number used in any calculation. We perform multiple random draws of synthetic source samples from this pool, calculate matrix elements for each draw, and then estimate the uncertainties. We begin with a small sample size, \( n = 100 \), and conduct 4096 random draws of this size from the full synthetic catalog, calculating matrix elements for each draw. The uncertainty for this sample size is then derived from the results by calculating the range that includes 68\% of the data around the mean. This procedure is repeated with increasing sample sizes ($n = 200, 400, … $), until we reach a size that is within 15\% of the size of the total pool (e.g., 10,000 for a pool of 11,500 sources). Each draw is made without replacement to avoid biases that might arise from reusing the same sources within a single estimation round. This stepwise approach allows us to observe how uncertainties in the recovery matrix elements vary with changes in the sample size (Figure \ref{fig:diag_el}), ensuring that the calculated uncertainties are representative of the true variability expected in different sampling scenarios.

\subsubsection{Forward modeling}

Numerous studies \citep[e.g.,][]{zwaan05, jones18} have demonstrated that the \himf can be effectively characterized by a \cite{schechter76} function, defined as
\begin{equation}
    \phi(M_\mathrm{HI}) = \ln(10) \: \phi_\ast \: \left( \frac{M_\mathrm{HI}}{M_\ast} \right)^{\alpha+1} \: e^{-\left( \frac{M_\mathrm{HI}}{M_\ast}\right)},
\end{equation}
whose three free parameters are the normalization constant \phistar, the ``knee" mass \mstar, and the low-mass slope $ \rm \alpha$. We adopt a forward modeling approach to estimate the Schechter function parameters that are consistent with our observed data. For each set of parameters ($\phi_\ast$, $M_\ast$, $\alpha$), we calculate the intrinsic number of \hi sources for each bin based on the volume of the survey and the bin widths. The intrinsic numbers for all bins (for a given set of Schechter function parameters) are then multiplied by the rows of the recovery matrix and summed across columns to determine the expected numbers of observed sources in all bins. The likelihood of observing the actual numbers of sources is then calculated based on these expected counts for each set of Schechter parameters during the fitting and MCMC sampling process. By directly estimating the Schechter function parameters using the observed set of detections, we avoid the need to create an unbiased (corrected, binned) histogram of detections for fitting with a Schechter function.

Our forward modeling approach allows us to convert the numbers of galaxies in the different \scisub{M}{inj} bins predicted for a given set of Schechter function parameters into numbers of recovered galaxies over the \scisub{M}{rec} bins. We calculate the total number of recovered galaxies in each \scisub{M}{rec} bin by summing the number of recovered galaxies from all the \scisub{M}{inj} bins. 
Our tests confirm that the inferred low-mass slope of the Schechter function can be biased if we fail to account for the recovery of sources with higher masses than the bins in which they were injected. Given that nontrivial numbers of sources are recovered with higher than injected masses, we have chosen to exclude the lowest \scisub{M}{rec} bin from the forward modeling process, as we cannot properly estimate the number of sources with $\log (M_{\rm HI}/M_\odot) < 7$ that have been recovered in that bin. 

At the high-mass end, our analysis includes the $10.5\leq \log(M_{\rm HI}/M_\odot)<11$ bin in which we actually detect no real \hi sources, in recognition of the fact that a non-negligible fraction of sources are recovered with higher masses and in order to determine the knee mass more accurately. Including the highest-mass bin in the forward modeling process constrains the number of sources in the second-highest mass bin to be consistent with the lack of detections in the highest mass bin. We note that the predicted number of sources in a given bin is the mean of a Poisson distribution describing the source count for that bin. The Poisson distribution $P(\mu)$ for mean values $\mu > 7$ can be approximated well with a Gaussian distribution $N(\mu,\sqrt{\mu})$. However, the discrepancy between the two distributions for smaller values of $\mu$ introduces a bias in the forward modeling process by differently weighting the estimated likelihoods for bins with few detections. We choose the Poisson likelihood, as our investigations reveal that choosing the Gaussian or Mean-Standard-Error (MSE) likelihood instead of the Poisson likelihood significantly underestimates the uncertainties associated with the inferred Schechter function parameters and introduces a bias towards steeper low-mass slopes.\footnote{For more details on these calculations, please see Appendix \ref{app:lkl}.}

\subsubsection{Ensuring consistency with the observed mass function}

It is necessary to maintain consistency between the observed \himf and the hypothetical \himf used in completeness calculations. For a given mass bin $i$, the completeness fraction $C_i$ represents the fraction of all sources in that bin that can be detected within the survey volume. In a traditional injection/recovery method, for bin $i$, the completeness fraction $C_i$ is defined as the ratio of the number of recovered sources $R_i$ to the number of injected sources $I_i$ in that bin, i.e. $C_i = R_i/I_i$. To account for cross-bin contamination, we can write
\begin{equation}
C_i = \frac{R_i}{I_i} = \frac{\sum_j I_j f_{j\rightarrow i}}{I_i} = \sum_j \frac{I_j}{I_i} f_{j\rightarrow i},
\end{equation}
where $f_{j\rightarrow i}$ represents the fraction of sources that are injected in bin $j$ and recovered in bin $i$. Isolating the term for galaxies both injected and recovered in bin $i$ results in
\begin{equation}
C_i = \frac{R_i}{I_i} = f_{i\rightarrow i} + \sum_{j\neq i} \frac{I_j}{I_i} f_{j\rightarrow i}, 
\end{equation}
showing that when some fraction of sources recovered in bin $i$ are from bin $j$ (for $j\neq i$), the completeness fraction for bin $i$ depends on the ratio of the number of injected galaxies in bin $j$ to bin $i$, i.e. $I_j/I_i$. In a traditional approach, the number of injections in each bin is proportional to the predicted count from the \himf to preserve relative ratios and enhance statistical reliability. For example, if the \himf predicts $N$ sources in a given bin, then the number of injected sources is some multiple of $N$. Given that any \himf varies significantly --- by more than two orders of magnitude --- across the mass range of a sample like ours, for every source in the highest mass bin, about 300 sources need to be injected into the lowest mass bin. This requirement poses a practical challenge for any traditional approach; for example, to obtain reliable statistics for our highest-mass bin if it were to contain only 100 sources, over 30,000 sources would need to be injected into the lowest-mass bin alone to maintain ratios consistent with the \himf. Using a recovery matrix, however, eliminates this constraint by incorporating cross-bin contamination directly into the matrix, allowing the number of injections in each bin to be chosen independently of the other bins. We make use of this flexibility when determining the number of injections by requiring a 10\% threshold on the uncertainties associated with the diagonal and below-diagonal matrix elements.

\subsubsection{Inferring Schechter function parameters}

In order to calculate the best-fit Schechter function parameters and their associated uncertainties, we first maximize the Poisson likelihood for our observed data to calculate a set of best-fit parameters, which are then used in a Markov chain Monte Carlo (MCMC) sampling process to estimate the uncertainties on those parameters. As discussed above, we inject enough sources in each \scisub{M}{inj} bin (independent from the other \scisub{M}{inj} bins) to estimate the diagonal and below-diagonal elements of the recovery fraction matrix with less than 10\% uncertainty. However, the estimated matrix elements and their uncertainties depend in part on the slope of the mass distribution of the injected sources {\it within} a given bin. A discrepancy between the observed \himf and the assumed slope of the mass distribution of sources in a given bin, if present, will result in a biased estimation of the Schechter function parameters. To mitigate this effect, we determine the best-fit Schechter function parameters using two iterations. In the first iteration, an initial recovery matrix is calculated using the ALFALFA $\alpha$.100 Schechter function parameters \citep{jones18}. This matrix is used to maximize the Poisson likelihood for our data to find an initial set of 
best-fit Schechter function parameters.
In the second iteration, we update the slope of the mass distribution of sources in each bin to be the slope recovered from the first iteration and recalculate the matrix elements, which are then used to find a revised set of
best-fit Schechter function parameters. 
The Schechter function parameters from the second iteration change by less than 1\% compared to the results from the first iteration, eliminating the need for any further iterations\footnote{We have confirmed that starting the iterations with a flat mass distribution across each bin similarly converges after two iterations, eliminating any concern that the use of an ALFALFA Schechter function to calculate the recovery matrix might somehow bias our results.}.

We use the PyMC\footnote{\url{https://www.pymc.io}} implementation of the MCMC method to estimate the uncertainties in the best-fit Schechter function parameters and their covariance by sampling an appropriate
posterior probability distribution. The prior distributions for the Schechter function parameters are log$_{10}$(\phistar) uniform in $[-4, -2]$, log$_{10}$(\mstar) uniform in $ [9, 10.5]$, and $\alpha$ uniform in $[-1.8,  -0.5]$. In the MCMC sampling process, we create 512 different realizations of the recovery matrix using the best-fit Schechter function parameters, in order to account for the uncertainties in the matrix elements, and perform an MCMC sampling with 8 chains and 32768 steps (after burn-in) for each of these 512 matrices. The posteriors from the 512 realizations are then combined to create a full posterior, for estimating the uncertainties on the best-fit Schechter function parameters in a way that includes the uncertainties in the matrix elements.

Taken together, the elements of the recovery matrix method for \himf\ determination --- starting with source injection and recovery as laid out in Section \ref{ssec:injec}, and ending with the estimation of Schechter function parameters as described immediately above --- could potentially be subjected to a full end-to-end validation test.  In such a test, ideally, a single realization of a mock galaxy population with a known \himf\ could be simulated into a source-free version of the LADUMA DR1 data cube, and the Schechter function parameters recovered from that mock population could be compared to those of the input \himf. Unfortunately, as discussed below in Section \ref{ssec:like} and Appendix \ref{app:lkl}, the appropriate (Poisson) uncertainties in the Schechter function parameters for a sample of the size being analyzed in this paper are very large, such that the parameters recovered for a single realization are likely to be very different from the input parameters {\it a priori}. As a result, (in)consistency between input and output \himf\ parameters for a single mock data cube cannot be used as a test of the recovery matrix method {\it per se}.  End-to-end validation of the method could in principle be achieved with an ensemble of mock data cubes; however, creating an ensemble of cubes that is sufficiently large to test the recovery matrix method (beyond the ability of Poisson uncertainties to compromise the test) would be computationally prohibitive and is beyond the scope of this paper.

\subsection{Modified maximum likelihood method} \label{ssec:mml}
In addition to the forward modeling approach discussed above, we have calculated the \himf using the modified maximum likelihood (MML) method described in \cite{obreschkow18}. The MML method, which assumes Poisson statistics, recovers the \himf without any binning while dealing with mass uncertainties and the selection biases present in the data. This method uses as an input a selection function that describes the recovery rate of sources as a function of distance and mass, which is averaged over extra variables such as width \citep{obreschkow18}. We estimate the selection function for our data cube by creating 2D histograms of the injected and recovered sources on a distance-mass grid. We have chosen 32 distance bins and 16 mass bins spanning the full distance and mass ranges of our data. The selection function is defined using a linear interpolator (in lieu of an analytical expression) on the 2D mass-distance gridded data, where the gridded values are calculated as ratios of the recovered source histogram and the injected source histogram. We use the {\tt dftools} package developed in R and described in detail in \cite{obreschkow18} to calculate the corrected number density of \hi sources in each bin. We then fit a Schechter function to the corrected number densities and estimate the uncertainties in the parameters of the Schechter function using MCMC sampling.

The MML method offers the flexibility to calculate the Schechter function parameters in several different ways, such as providing the maximum volume $V_{\rm max}$ in which each source can be detected, or alternatively, providing a selection function\footnote{For more details, see \url{https://rdrr.io/github/obreschkow/dftools/man/dffit.html}.}. While using the maximum volume option results in a reasonable fit to our data, we chose to generate a selection function based on our extensive injection/recovery tests for a fair comparison with the recovery matrix method, as described in Section \ref{ssec:mml}.\footnote{To allow for a fair comparison between the results of the MML and RM methods, we need to use the selection function approach for the former, so that we can make use of the same injection/recovery information used by the latter.} However, this approach results in convergence issues with the MML method, resulting in an unreliable fit. To work around this problem, we use the built-in functionality of the {\tt dftools} package to extract the corrected \himf values for our mass bins and subsequently fit a Schechter function to these values. This process closely aligns with previous \himf studies, where the corrected \himf is calculated from binned data and a Schechter function is then fit to those values, making this discussion relevant to previous works that follow a similar analytical framework.

\begin{figure*}
\centering
\includegraphics[width=0.7\textwidth]{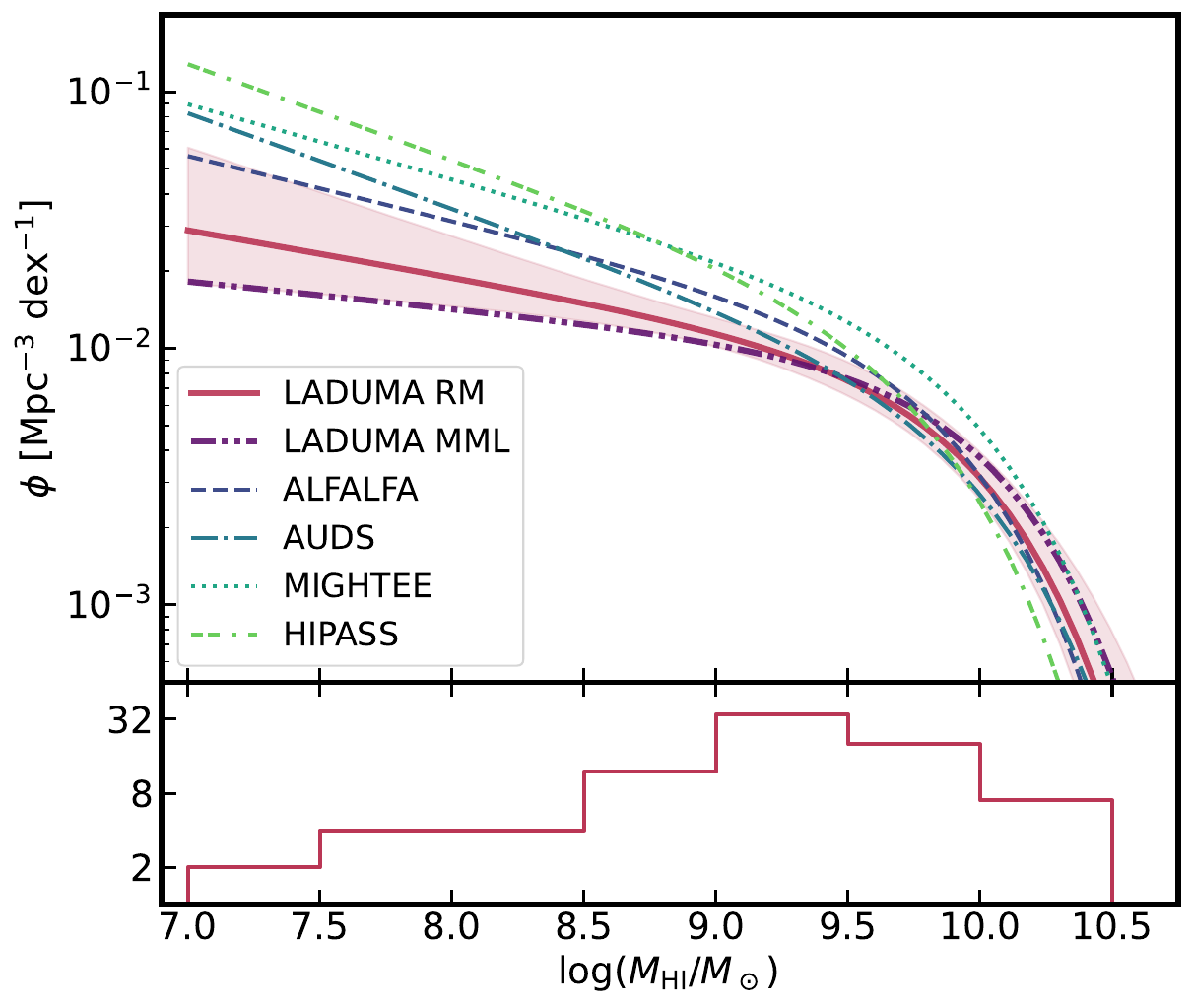}
\caption{The \himf for LADUMA as derived via the recovery matrix and the MML methods in comparison with previous measurements in the literature. The best-fit Schechter functions based on the recovery matrix (RM) and the MML methods are shown with the solid and dash-double-dot lines. The shaded region represents the 1$\sigma$ uncertainty in the RM fit, sampled from the MCMC posteriors (Figure \ref{fig:corner}); the analogous uncertainty in the MML fit is shown in Figure \ref{fig:mml}. The histogram in the lower panel shows the number of observed sources in the LADUMA high-purity sample as a function of \hi mass. The lowest-mass bin shown here (with two detections) was not included in the forward modeling and posterior sampling (see Section \ref{sec:disc}.).
\label{fig:sch_comp}}
\end{figure*}

\section{Results} \label{sec:res}
\subsection{\himf and Schechter function parameters} \label{subsec:params}

\begin{table*}[h]
\caption{The best-fit parameters of the Schechter function and corresponding galactic \ohi values for LADUMA (this work), ALFALFA, HIPASS, AUDS, and MIGHTEE. $z_{\rm max}$ shows the upper limit of the redshift range for each survey. For LADUMA, 
the \textit{disfavored} MML ``free fit'' shows the best-fit values obtained by excluding the highest-mass bin, which has no detections, from the fitting process. We calculate the comoving volumes using the area and redshift range reported for each survey in the corresponding publication, scaled as needed to our cosmology. All other parameters are rescaled as needed for $H_0 = 70 $ km~s$^{-1}$~Mpc$^{-1}$.}
\label{tab:schech}
% \tabletypesize{8pt}
{\centering
\movetableright=-1in
\begin{tabular}{lccccccc}
\hline
Survey &  $z_{\rm max}$ & Volume (Mpc$^{3}$) & $N_{\rm src}$& $\phi_\ast$ ($\times 10^{-3}$ Mpc$^{-3}$) & log($M_\ast/M_\odot$) & $\alpha$  & \ohi ($\times 10^{-4}$) \\
\hline
LADUMA RM & $0.088$ & $\sim 9\times 10^3 $& 82 &$3.56_{-1.92}^{+0.97}$ & $10.01_{-0.12}^{+0.31}$ & $-1.18_{-0.19}^{+0.08}$ & $3.09_{-0.47}^{+0.65}$ \\
LADUMA MML &  $''$ &  $''$ & $''$ & $3.78_{-1.14}^{+1.81}$ & $10.07_{-0.17}^{+0.14}$ & $-1.10_{-0.12}^{+0.16}$ & $3.48_{-0.93}^{+0.26}$ \\ 
\textit{LADUMA MML}&  $\mathit{''}$ &  $''$ & $''$ & $\mathit{3.61_{-2.29}^{+0.47}}$ & $\mathit{10.10_{-0.09}^{+0.41}}$ & $\mathit{-1.12_{-0.22}^{+0.05}}$ & $\mathit{3.62_{-0.97}^{+0.67}}$ \\
\textit{(free fit)}\\
ALFALFA (J18)& $0.05$& $\sim 6.2 \times 10^6$\tablenotemark{\rm \textdaggerdbl} & $22831$ & $4.50_{-0.82}^{+0.82}$ & $9.94_{-0.05}^{+0.05}$ & $-1.25_{-0.10}^{+0.10}$ & $3.90_{-0.61}^{+0.61}$\tablenotemark{\rm \dag} \\
HIPASS (Z05) & $0.042$& $\sim 12\times 10^6$ & $4315$ & $4.88_{-0.81}^{+0.81}$ & $9.86_{-0.04}^{+0.04}$ & $-1.37_{-0.06}^{+0.06}$ & $3.75_{-0.61}^{+0.61}$ \\
AUDS (X21) & $0.16$& $\sim 39\times 10^3$ & $247$ & $2.65_{-0.48}^{+0.48}$ & $10.06_{-0.04}^{+0.04}$ & $-1.37_{-0.03}^{+0.03}$ & $3.33_{-0.10}^{+0.10}$ \\
MIGHTEE (P23) & $0.084$ & $\sim 21\times 10^3$\tablenotemark{$*$} & $203$ &$5.12_{-2.89}^{+5.58}$ & $10.04_{-0.24}^{+0.24}$ & $-1.29_{-0.26}^{+0.37}$ & $5.26_{-0.95}^{+0.91}$  \\
\hline
\end{tabular}}
\tablenotetext{\text{\textdaggerdbl}}{
We list here the comoving volume probed by the ALFALFA $\alpha$.100 sample for our assumed cosmology, which is slightly smaller than the $6.5 \times 10^6\,{\rm Mpc}^3$ actually used in the \himf calculations of \citet{jones18} (M. Jones, private communication). Use of our smaller volume would modestly increase $\phi_\ast$ and \ohi for ALFALFA relative to the values reported in this table.}
\tablenotetext{\dag}{The reported ALFALFA \ohi is corrected for \hi self-absorption; the value without this correction is $3.50_{-0.51}^{+0.51}$.}
\tablenotetext{$*$}{We list here the comoving volume probed by the MIGHTEE analysis of the COSMOS and XMM-LSS fields for our assumed cosmology, which is (a) larger than the $7 \times 10^3\,{\rm Mpc}^3$ ``cosmological volume'' reported in \citet{pono23} that includes an unnecessary extra factor of $h^{-3}$, but (b) only slightly 
different from the volume actually used in that paper's $V_{\rm eff}$ analysis (A. Ponomareva, private communication).  Use of our 
slightly different survey volume in the context of a $V_{\rm eff}$ analysis would modestly 
change $\phi_\ast$ and \ohi for MIGHTEE relative to the values reported in this table.
}
\tablenotetext{}{References in the table are as follows: J18= \cite{jones18}, Z05= \cite{zwaan05}, X21= \cite{xi21}, and P23= \cite{pono23}.}
\end{table*}

\begin{figure}
\centering
\includegraphics[width=0.45\textwidth]{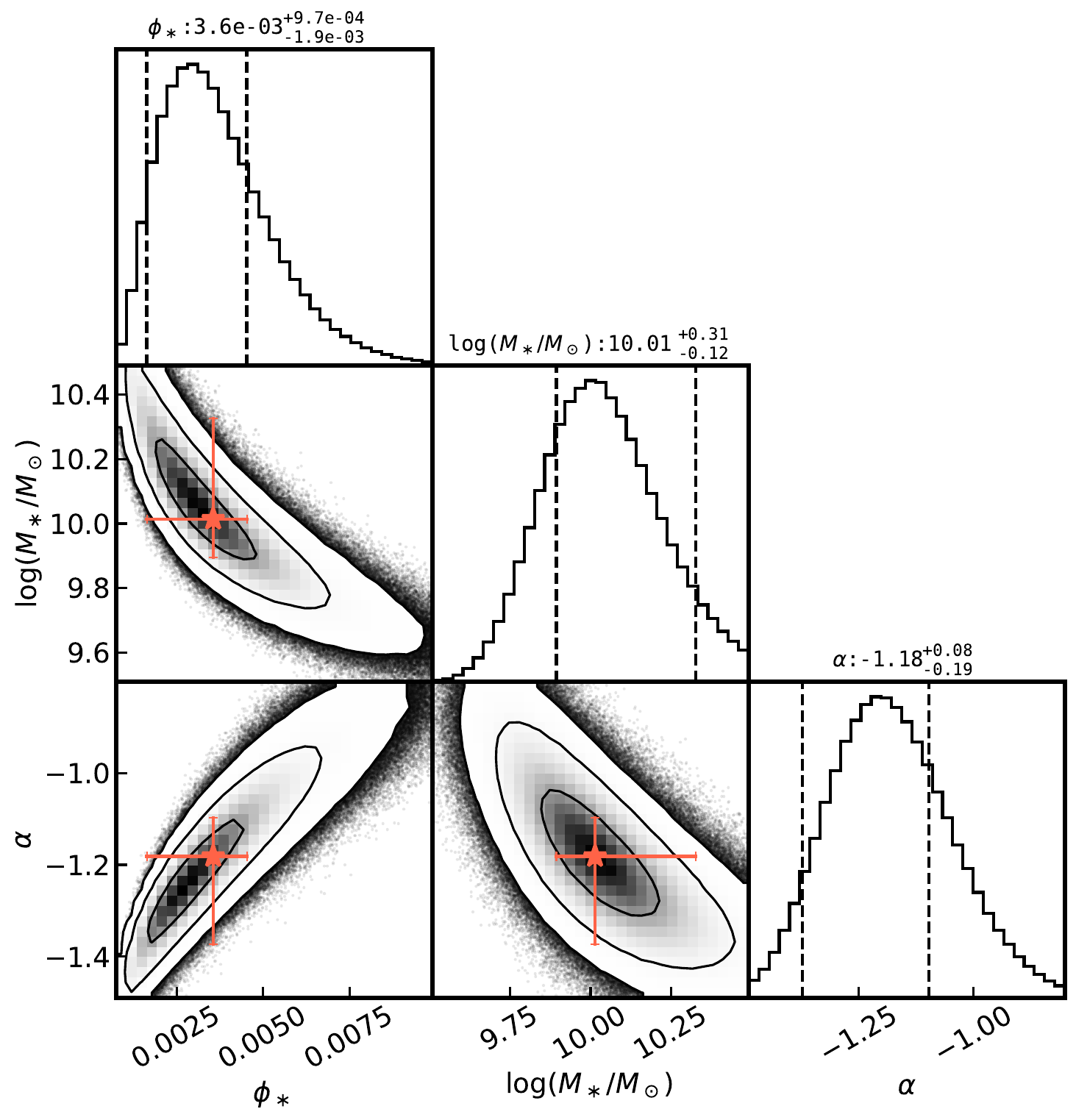}
\caption{Marginalized 1D and 2D distributions of the sampled posterior for the recovery matrix method. The best fit values and their associated uncertainties, obtained from the 3D posterior, are shown on the 2D marginalized distributions. The vertical dashed lines on the 1D histograms show the 1$\sigma$ uncertainty regions for each parameter. The contours show $\{1,2,3\}\sigma$ level (i.e., 39th, 86th, and 99th percentile) uncertainties for the 2D marginalized distributions.
\label{fig:corner}}
\end{figure}

In Figure \ref{fig:sch_comp}, we compare the Schechter functions preferred by the recovery matrix method and the MML method for the LADUMA DR1 data, along with the actual detected number of galaxies in each bin within our high-purity sample. The best fit parameters for both methods are determined using a search on a variable resolution grid with a high density of points around the best-fit region. Figure \ref{fig:corner} shows the estimated Schechter function parameters and their uncertainties plotted on the marginalized 1D and 2D posterior probability distributions for the recovery matrix method. The associated uncertainties in the parameters are determined directly from the 3D sampled posterior, as none of the three Schechter function parameters are considered nuisance parameters.\footnote{The 4D volume enclosed inside the 1$\sigma$ limit of a 3D Gaussian distribution is 24.91\% of the total volume. Estimating the parameters and their uncertainties from the marginalized posterior distributions does not lead to the same results as when parameters are estimated from the 3D distribution (see Figure \ref{fig:corner}).}. 

\begin{figure}
\centering
\includegraphics[width=0.45\textwidth]{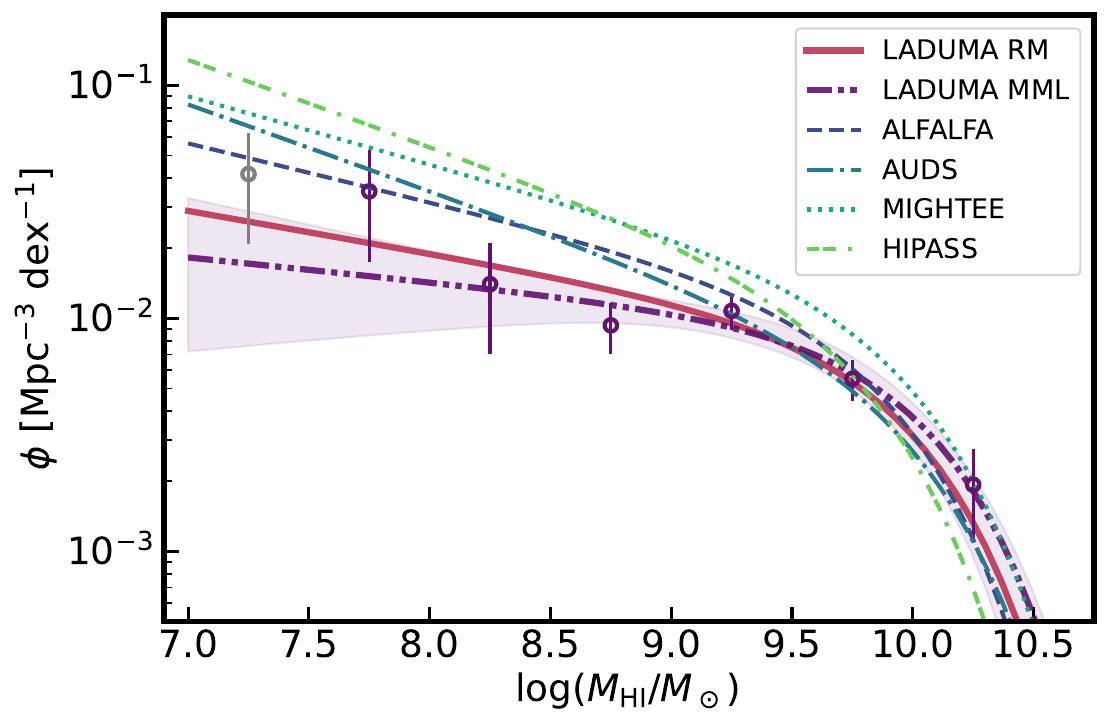}
\caption{The \himf for LADUMA from the MML method (see Section \ref{ssec:mml}) in comparison with previous measurements in the literature. The best-fit Schechter function is shown as the solid line; the shaded region represents the 1$\sigma$ uncertainty range on the MML fit. The lowest mass bin datapoint, shown in grey, is not included in the fitting process.
\label{fig:mml}}
\end{figure}

Determining the appropriate uncertainties for the MML result is complicated by the fact that the MML method does not provide uncertainties for any bin with zero detections. If we circumvent this limitation by excluding the zero-detection highest-mass bin, we obtain only a weak constraint on the knee mass, with a 1$\sigma$ upper limit reaching the upper limit ($\log (M_\ast/M_\odot) = 10.5$) of the MCMC prior range. Excluding the highest mass bin from the fitting process eliminates any loss associated with the separation between the Schechter function value for that bin and its observed value of zero, essentially allowing the fit to extend more freely to inappropriately large values at the high-mass end. In order to improve the Schechter function fit to the MML results, we assign an artificial value and an associated uncertainty to the highest-mass bin when we fit the Schechter function and perform MCMC sampling. As the knee mass and its associated uncertainties are sensitive to the artificial value assigned to the highest mass bin ($10.5\leq \log(M_{\rm HI}/M_\odot)<11$), we assign a value of 0.4 detections in that bin (only modestly higher than the actual value of 0, and encompassing both 0 and 1 within its $1\sigma$ Poisson uncertainty). 
This adjustment to the highest-mass bin leads to more reliable estimates of the knee mass and its associated uncertainties. Given that the output of the MML method is the corrected number density and associated uncertainty for each bin, we need to use a Gaussian likelihood for fitting and MCMC sampling. Therefore, the MML results are slightly biased by the effects of bins with low number of counts (as explained in Appendix \ref{app:lkl}) compared to the results from the recovery matrix method. The results from the MML method, which are consistent with our recovery matrix method results but less so with the results of previous surveys, are shown in Figure \ref{fig:mml}. 

\begin{figure}
\centering
\includegraphics[width=0.45\textwidth]{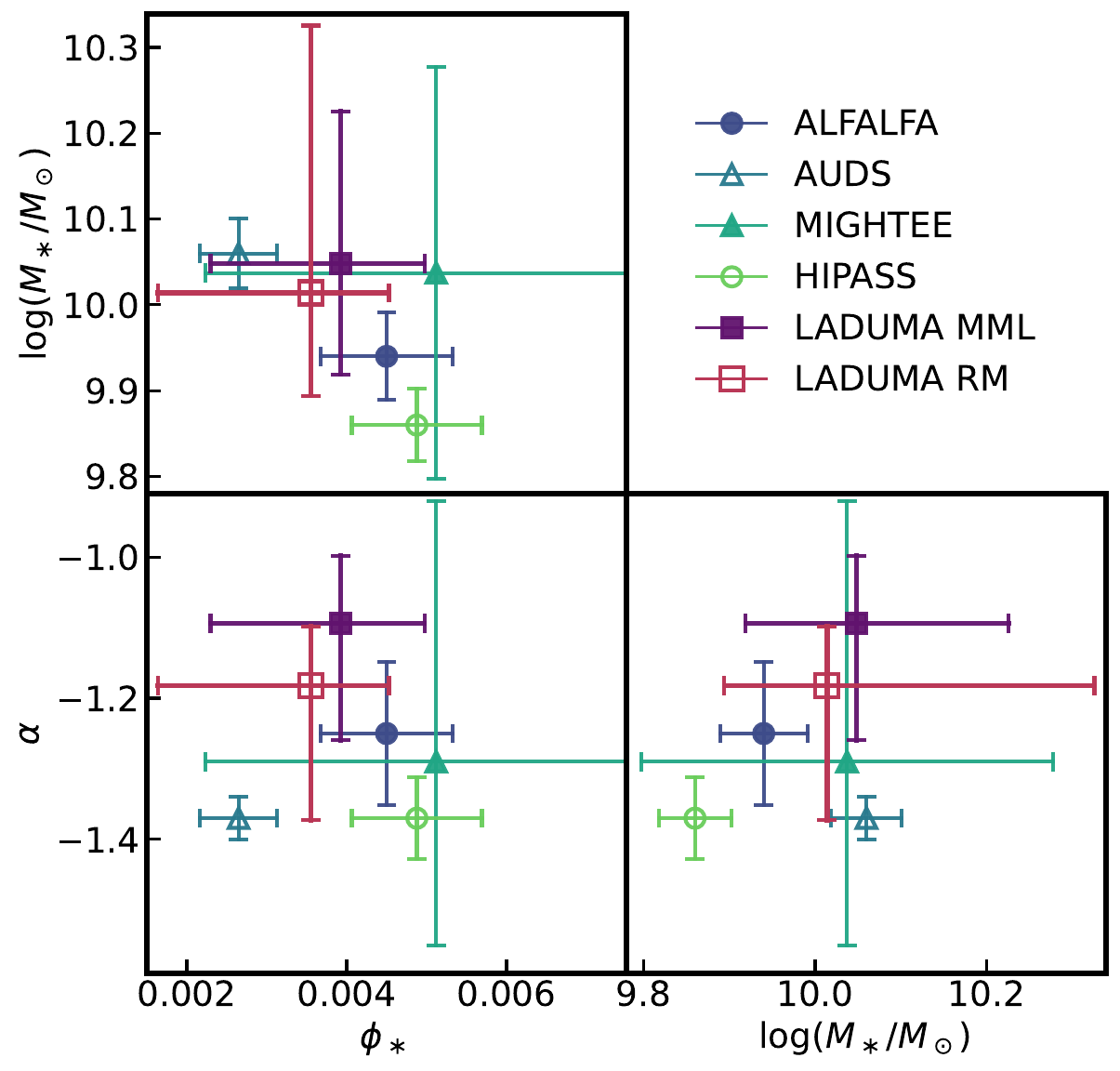}
\caption{Schechter function parameters with associated uncertainties for different surveys. Square symbols represent LADUMA measurements. LADUMA's $\phi_\ast$ and $M_\ast$ agree with the results for all the previous surveys within 1$\sigma$ uncertainties.  LADUMA's low-mass slope is shallower than those measured in all previous surveys, although it is in agreement within the 1$\sigma$ uncertainty with ALFALFA and MIGHTEE results. The values for these parameters are reported in Table \ref{tab:schech}, along with the literature references for previous surveys. The best-fit parameters from previous surveys have been rescaled to $H_0$ = 70 km~s$^{-1}$~Mpc$^{-1}$.}
\label{fig:sur_comp}
\end{figure}

Comparing our results with those of previous surveys is 
complicated by the fact that different authors have used different values of 
$H_0$ (with which \phistar, \mstar, and \ohi scale straightforwardly),  
different approaches to defining cosmological volume (which are not always stated
but will affect \phistar and \ohi), and different choices of likelihood that translate to different ``$1\sigma$'' uncertainties.  To provide as consistent comparisons with 
previous results as we can, in Table \ref{tab:schech} we have scaled the values 
of \phistar, \mstar, and \ohi reported by the HIPASS, AUDS, and MIGHTEE teams for our choice of $H_0$ (ALFALFA uses the same value of $H_0$, so requires no 
rescaling), and we provide indications in the table notes of how we might 
expect results to change further on the basis of volume calculations. We also 
present the best-fit Schechter function parameters for the recovery matrix method and the MML method for LADUMA, along with a ``free fit'' version of the MML results that illustrates the impact of ignoring the zero-detection highest-mass bin. Figure \ref{fig:sur_comp} shows that the estimated Schechter function parameters from the recovery matrix and MML methods are in good agreement with each other.
Relative to previous \himf measurements from the literature, while the individual Schechter function parameters that we recover with the MML method are not consistent with all previous results (e.g., the AUDS measurement for $\alpha$), those
we recover for LADUMA using our (preferred) recovery matrix method agree with those for all previous surveys within the associated $1\sigma$ uncertainties when added in quadrature. The situation for the \himf as a whole is not as clear: due to 
covariance between the Schechter function parameters, some of the literature H{\sc{i}}MFs fall outside the $1\sigma$ uncertainty swath shown in Figure \ref{fig:sch_comp} across a wide range of masses.
A full assessment of the consistency of our derived \himf with the results of previous surveys, which would require detailed knowledge of the covariances among those surveys' respective Schechter function parameters and perhaps recalculation of their uncertainties using a Poisson likelihood (which then might or might not overlap with the LADUMA uncertainty range), is beyond the scope of this paper.

\subsection{Cosmic \hi density (\ohi)}

A complete inventory of \hi\ in the local universe would 
include neutral hydrogen that lies outside of galaxies, 
both within and beyond the cosmic web. In this paper, we 
can estimate the contribution of galaxies to
\ohi  
based on the comoving \hi mass density ($\rho_{\rm HI}$) that is calculated by integrating the Schechter function. The density $\rho_{\rm HI}$ is computed analytically \citep{meyer17} as 
\begin{equation}
\rho_{\rm HI} = \Gamma(\alpha+2) \phi_\ast M_\ast, 
\end{equation}
where $\Gamma$ is the Euler gamma function and $\phi_\ast$, $M_\ast$, and $\alpha$ are the Schechter function parameters. The contribution of galaxies to \ohi is then calculated as 
\begin{equation} \label{eq:ohi}
\Omega_{\rm HI} = \frac{8 \pi G}{3 H_0^2}~ \rho_{\rm HI}, 
\end{equation}
where $G$ is the gravitational constant and $H_0$ is the Hubble constant. We measure a galactic \ohi = $3.09_{-0.49}^{+0.58} \times 10^{-4}$ with the recovery matrix method and a galactic \ohi = $3.48_{-0.93}^{+0.26} \times 10^{-4}$ with the MML method. Uncertainties in 
these measurements 
encompass 68\% of \ohi values (around the best-fit value) calculated from 4096 sets of Schechter function parameters randomly drawn from the 3D sampled posterior. Our results are in agreement with the results for all the previous surveys except for MIGHTEE within 1$\sigma$ uncertainties added in quadrature.

\section{Discussion} \label{sec:disc}

\begin{figure}
\centering
\includegraphics[width=0.45\textwidth]{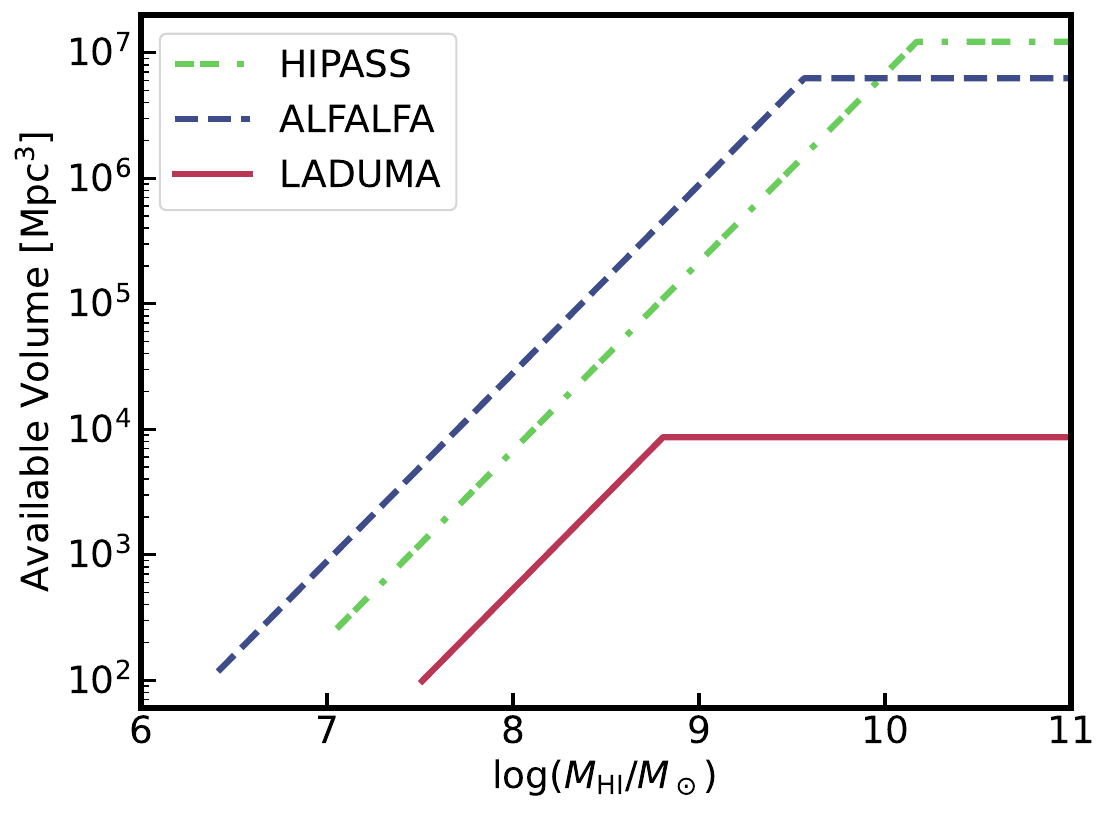}
\caption{Volumes accessible for the detection of sources with given $M_{\rm H\,I}$ by LADUMA, ALFALFA, and HIPASS, calculated based on their respective detection limits. The maximum volume inside which \textit{all} sources of \hi mass $M_{\rm H\,I}$ are detected cannot be precisely defined, as it depends on parameters such as velocity width. However, we can roughly define an ``accessible'' volume beyond which a given survey in practice detects vanishingly few sources of \hi mass $M_{\rm H\,I}$; these are the volumes plotted above. The number density distribution for sources of \hi mass $M_{\rm H\,I}$ is sensitive to the distribution of sources (large scale structure) inside the corresponding ``accessible" volume. While all surveys have a much smaller (by several orders of magnitudes) available volume at lower masses compared to higher masses, that discrepancy is less significant for LADUMA compared to HIPASS and ALFALFA. The fact that the flat part of the LADUMA curve (corresponding to its {\it total} survey volume) extends farther left than do the flat parts of the HIPASS and ALFALFA curves is due to LADUMA's greater depth (and in spite of LADUMA's slightly larger redshift range) relative to the other two surveys.}
\label{fig:volume_comp}
\end{figure}

\subsection{Available volume as a function of \hi mass} \label{ssec:comp}

The \himf we have measured from the LADUMA data is subject to some of the same 
caveats as mass functions measured from previous surveys, in ways that are
instructive about the limitations that apply to any \himf\ 
measurement. One significant caveat is that the 
detectability of a source depends on its \hi\ velocity width 
as well as its \hi\ mass; our recovery matrix approach deals 
with this complication by marginalizing over the expected 
distribution of disk inclinations 
\citep[see, e.g.,][]{obreschkow18}. Perhaps 
the most important caveat is that 
for a flux-limited \hi sample, the sources in the lowest-mass bin are typically confined to a much smaller (nearby) volume than the total survey volume \citep[][see especially their Figure 4]{Zwaan97}. 
While previous authors have emphasized the importance of the bias of a high-mass galaxy population relative to the underlying dark matter distribution as a contributor to cosmic variance \citep[e.g.,][]{moster11}, the limited volume in which low-mass galaxies can be detected for a flux-limited sample can be just as important. Among the three Schechter function parameters, $M_\ast$ will be least sensitive to ``available'' volume and most sensitive to population bias, 
as it is predominantly determined by the high-mass bins. 
In contrast, and as pointed out by \citet{jones18}, $\alpha$ is expected to be most sensitive to available volume (albeit least sensitive to population bias), as it is predominantly determined by the low-mass bins. 
$\phi_\ast$ is expected to fall somewhere in between in terms of susceptibility to cosmic variance, as all bins contribute to the overall normalization within the limits of their associated uncertainties. For our data, the accessible volume associated with the lowest-mass bin, in which we recover any sources in our high-purity sample (specifically, two sources), is only about 2\% of the total survey volume. 
We find that including this bin in our calculations for the best-fit Schechter function parameters would result in a significantly steeper value for $\alpha$. Our decision to exclude this bin from our calculations reduces the 
impact of the extremely small volume of the lowest mass bin. Appendix \ref{app:unc} uses the ALFALFA $\alpha.100$ galaxy catalog to demonstrate that cosmic variance increases the uncertainty 
in the number of high-mass (${\rm log}\,(M_{\rm HI}/M_\odot) \geq 9.75$) galaxies in the LADUMA volume compared to Poisson uncertainties alone, but the accessible volume subtleties discussed above preclude any extension of this analysis to individual Schechter function parameters.

\begin{figure}
\centering
\includegraphics[width=0.45\textwidth]{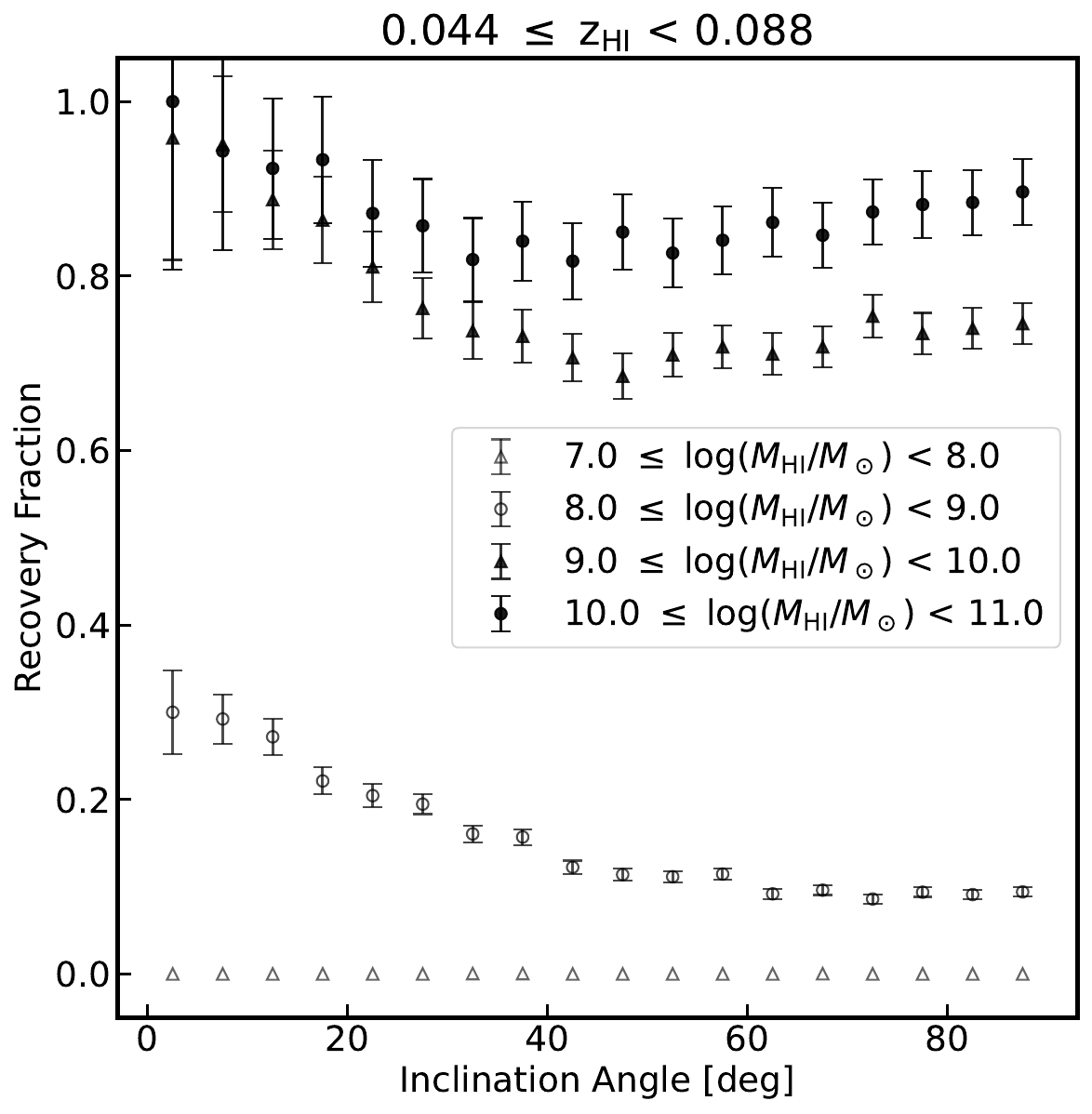}
\caption{Recovery fraction as a function of source inclination for the higher-redshift half of LADUMA's low-$z$ SPW. While the recovery fraction for low-mass sources decreases significantly at high inclinations, the recovery fraction for high-mass sources remains consistently high at all inclinations.}
\label{fig:inc_rec}
\end{figure}

We note that LADUMA does derive some benefit for \himf derivation from its greater depth compared to previous surveys: as Figure \ref{fig:volume_comp} illustrates, the volume that is available to LADUMA for constraining the \himf at ${\rm log}\,(M_{\rm H\,I}/M_\odot) \leq 8.5$ is only $\sim 1$ order of magnitude smaller than those that were available to HIPASS and ALFALFA, in contrast to the $\sim 4$ orders of magnitude advantage those surveys have over LADUMA in terms of available volume for the highest (${\rm log}\,(M_{\rm H\,I}/M_\odot) \geq 10$) masses.  From this point of view, the fact that LADUMA recovers a faint-end slope $\alpha$ flatter than (but consistent within the errors with) ALFALFA --- and indeed matching the ALFALFA measurements of $\alpha$ in 
certain of that survey's sub-volumes \citep{jones18} --- is not an unexpected result.  At higher masses, LADUMA benefits in a different way from its greater depth, namely superior resilience against the systematic loss of high-inclination, high-mass sources at large distances within its observed sample, as demonstrated in Figure \ref{fig:inc_rec}. 

\subsection{More complex models}

The behavior of the observed LADUMA \himf (see Figure \ref{fig:mml}), particularly the rolloff just below $M_\ast$ followed by an upturn at lower masses, is reminiscent of features seen in optical/near-IR luminosity functions (LFs). These LFs can be effectively decomposed by galaxy type or color into multiple LFs, each represented by its own Schechter function \citep{Sandage85, Loveday12, Driver22}. Previous studies \citep[e.g.,][]{said19, jones20} have shown a dependency of the Schechter function parameters on environment, suggesting that a more complex functional form (e.g., different Schechter functions for high-density vs. low-density environments) might be required to describe the overall \himf accurately. Our forward-modeling approach is designed to accommodate various forms of mass functions beyond the traditional Schechter form, providing a flexible framework for analyzing the \himf in greater detail for larger datasets. In addition to more complex functional forms, our approach can easily incorporate non-parametric models by directly sampling the intrinsic numbers of galaxies in different bins based on different priors, predicting corrected number densities for all bins. However, our current dataset does not permit the fitting of models with more degrees of freedom, as the relatively small number of sources available cannot adequately constrain a more complex model.

\subsection{Determining completeness} \label{ssec:correction}
 
This subsection and the next 
highlight two challenges in \himf determination, illustrating the merits and limitations of the recovery matrix approach compared to traditional methods. The completeness of a survey for sources of a certain mass can be assessed through two primary methods. The effective volume ($V_{\rm eff}$) method estimates the survey's completeness for sources of mass $M$ by evaluating the effective volume within which these sources can be observed\footnote{The effective volume ($V_{\rm eff}$) for sources of mass $M$ is equal to the harmonic mean of the maximum volume ($V_{\rm max}$) --- calculated by accounting for the dimming of the signal as the distance to the source increases and taking into account the sensitivity of the survey as a function of distance and position --- for each source with mass $M$ detected in the survey \citep{obreschkow18}. $V_{\rm max}$ is a function of \mhi\ and \hi\ velocity width ($\Delta v_{\rm HI}$), while $V_{\rm eff}$ is a function of \mhi\ only.}. Injection/recovery methods offer an alternative approach by introducing synthetic sources into the datacube and tracking their detection or non-detection using the same processes applied to real sources. These approaches can be encapsulated in expressions for the measured number density of sources of mass $M$, $\phi(M)$,
\begin{align}
    \phi(M)\,d{\rm log}\,M & = \frac{n_o(M)}{C_s(M) V_s} &\text{ (Injection/recovery)}\label{eq5:injrec} \\
    &= \frac{n_o(M)}{V_{\rm eff}(M)}, &\text{ (Effective volume)}\label{eq5:veff}
\end{align}
where $n_o(M)$ represents the number of sources detected in a bin of width $d{\rm log}\,M$ centered at $M$, $V_{\rm eff}(M)$ is the effective volume for sources with \hi mass $M$, $V_s$ is the total volume of the survey, and $C_s(M)$ is the completeness of the survey for sources with \hi mass $M$ (estimated via injection/recovery tests). By comparing equations \ref{eq5:injrec} and \ref{eq5:veff}, it becomes evident that the effective volume is equivalent to the completeness multiplied by the total volume of the survey. The completeness $C_s$ can be further decomposed as $C_s(M) = C_p(M)\cdot C_V(M)$, where $C_p(M)$ is the completeness based on the distribution of source properties, such as inclination, and $C_V(M)$ is the fraction of the total volume of the survey available for detection of sources of mass $M$. As described in Section \ref{ssec:comp} and demonstrated in Figure \ref{fig:volume_comp}, the accessible volume for sources of mass $M$ can vary drastically, e.g., by orders of magnitude, over the mass range of a survey. The completeness for low-mass sources is largely determined by the fraction of the total volume accessible to those sources and less so by the distribution of their intrinsic properties. The combination of these factors reveals that fluctuations (e.g., in the number of observed sources, which is sensitive to the distribution of the sources) in the small volume available to sources at the low-mass end of the mass range are amplified by the inverse of the accessible volume fraction in the completeness correction process. Therefore, it is essential to limit the amplification of these fluctuations, which can be done by imposing a minimum accepted volume fraction (10\% for this work) for the low vs. the high end of the mass range.

Beyond the amplification by the inverse of the volume fraction, as laid out in detail in Appendix \ref{app:veff}, the accuracy of the completeness correction in the $V_{\rm eff}$ method is sensitive to the distribution of properties in the observed population. To demonstrate this point, we consider two different types\footnote{Type can represent any property of a source that affects the volume within which it can be detected; one example would be inclination.} of sources of mass $M$, face-on and edge-on. For unresolved detections, it is clear that edge-on sources can only be detected in a smaller volume than otherwise identical face-on sources, since their larger line-of-sight velocity widths translate to lower peak flux densities. Let us represent the ratio of the volume available to face-on sources to the volume available to edge-on source as $R_{f/e}$. Assuming (unrealistically) that there are intrinsically the same numbers of edge-on and face-on sources in a survey volume, we would expect that for each edge-on source we detect, $R_{f/e}$ face-on sources should be detected. As described in Appendix \ref{app:veff}, the effective volume calculated for sources of mass $M$ is sensitive to the \textit{observed} ratio of the number of face-on to the number of edge-on sources. However, for an injection/recovery approach, this sensitivity would not affect the inferred completeness of the survey as long as the underlying distribution of sources is well understood\footnote{It is noted that for an injection/recovery method any bias in the assumed underlying distributions of source parameters would result in a similar effect and introduce additional uncertainty in the compeletness fractions.}, since the completeness is determined through source injection. 

\subsection{Choice of likelihood} \label{ssec:like}

Another challenge
for \himf determination 
involves accurately addressing uncertainties in the fitting process, particularly in bins with few or no detections. The data consist of observed source counts across various mass bins, each representing a count randomly drawn from a Poisson distribution. A traditional approach assumes that the counts are Gaussian-distributed with a Poisson uncertainty --- computed as the square root of the number of sources in each bin --- and uses a completeness correction factor to calculate the corrected number density of \hi sources per bin. These values are then used to estimate the best-fit Schechter function parameters and their uncertainties. As outlined in Appendix \ref{app:lkl}, using a Gaussian approximation for the full Poisson likelihood significantly biases the derived best-fit parameters of the Schechter function and tends to underestimate the 1$\sigma$ uncertainties, especially in bins with few detections. This underestimate affects both bins where completeness is high but galaxies are few in number, i.e., at high masses, and bins where galaxies are plentiful but completeness is low, i.e., at low masses.\footnote{In practice, surveys with smaller volumes and fewer total detections can mitigate this problem by using coarser mass bins, each containing a larger number of sources, when inferring \himf\ parameters.} The root of this discrepancy lies in the asymmetric and positively skewed nature of the Poisson distribution. For a bin with a small number of detections, assuming Gaussian-distributed counts skews the estimated mean toward the observed count, leading to over-fitting of the data. 
For LADUMA, this bias would manifest as an unrealistically steep low-mass 
slope, since (as is typical for any survey) the lowest-mass bins contain 
relatively few detections.
In contrast to a traditional approach, the recovery matrix method employs forward modeling to predict the mean of the underlying Poisson distribution for each mass bin for any set of Schechter function parameters and implements a Poisson likelihood for model fitting and MCMC sampling. One significant advantage of forward modeling is its ability to integrate bins with zero detections into the fitting process, enhancing the robustness of the fit. As discussed in \ref{subsec:params}, for the MML method, we observe significantly larger uncertainties in the knee mass parameter when we exclude the zero-detection highest-mass bin.  However, forward modeling naturally accommodates such cases, as it models uncertainties based on the predicted values and is compatible with the assumption of a Poisson distribution for the observed counts.

\section{Summary and Conclusions} \label{sec:sum}

In this paper, we present a comprehensive analysis of the \himf and the contribution of galaxies to \ohi from a portion of the Looking At the Distant Universe with MeerKAT Array (LADUMA) survey covering $0\leq z \leq 0.088$. Using a high-purity sample of \hi detections from LADUMA's Data Release 1, we develop a new ``recovery matrix" method and benchmark it against a traditional maximum likelihood approach for measuring the \himf. Our estimates of the Schechter function parameters are $\phi_\ast = 3.56_{-1.51}^{+1.79}\times 10^{-3}$ Mpc$^{-3}$ dex$^{-1}$, $\alpha = -1.18_{-0.14}^{+0.14}$, and $\log(M_\ast/M_\odot) = 10.01_{-0.17}^{+0.23}$. These values are in agreement with previous measurements of the \himf over similar redshift ranges, enhancing confidence in the robustness of our results and the resilience of LADUMA's single-pointing geometry against the effects of cosmic variance.

Our methodology uses extensive synthetic source injections to correct for survey incompleteness and, for the first time, includes the effects of the continuum subtraction process. We account for varying sensitivity and completeness across the survey volume and mass range. In particular, our forward modeling approach proves beneficial in handling bins with few or no detections (using an appropriate Poisson likelihood during the fitting and MCMC sampling process), thereby minimizing systematic effects on the derived Schechter function parameters. By using a Poisson likelihood instead of a Gaussian likelihood with Poisson uncertainties, we avoid overfitting the low-mass and high-mass bins of the Schechter function, which usually have few detections. Our analysis highlights the importance of cultivating a high-purity sample for reliably estimating survey completeness and avoiding biases in the recovery of mass function parameters. The recovery matrix allows for independent determination of the required number of injections per mass bin, ensuring reliable statistics across all bins --- overcoming a challenge faced by traditional injection/recovery methods. 

Looking forward, the LADUMA survey's future data releases will allow us to refine these measurements and potentially reveal new aspects of \hi evolution across a broader redshift range (e.g., Hoosain et al., in preparation). The continued development and application of new methods like the recovery matrix will be crucial in leveraging the full potential of these data to enhance our understanding of galaxies' \hi distributions and their role in galaxy evolution.

%% The "ht!" tells LaTeX to put the figure "here" first, at the "top" next
%% and to override the normal way of calculating a float position
%% Also note that the akcnowlodgment environment does not support long amounts of text. If you have a lot of people and institutions to acknowledge, do not use this command. Instead, create a new \section{Acknowledgments}.
\section{Acknowledgments} \label{sec:ack}

The MeerKAT telescope is operated by the South African Radio Astronomy Observatory (SARAO; \url{www.sarao.ac.za}), which is a facility of the National Research Foundation (NRF), an agency of the Department of Science and Innovation. The authors thank the members of the SARAO engineering, commissioning, and science data processing teams for building and operating an absolutely superb facility. The MeerKAT data presented in this paper were processed using the ilifu cloud computing facility (\url{www.ilifu.ac.za}), a partnership between the University of Cape Town (UCT), the University of the Western Cape (UWC), the University of Stellenbosch, Sol Plaatje University, and the Cape Peninsula University of Technology. The ilifu facility is supported by contributions from the Inter-University Institute for Data Intensive Astronomy (IDIA, which is a partnership between the University of Cape Town, the University of Pretoria and the University of the Western Cape), the Computational Biology division at UCT, and the Data Intensive Research Initiative of South Africa (DIRISA). Data processing used pipelines that were developed at IDIA and are available at \url{https://idia-pipelines.github.io}.

The authors thank the anonymous referee for a constructive and helpful report and Matt Bershady, Barbara Catinella, Munira Hoosain, Ren\'ee Kraan-Korteweg, and Dominik Schwarz for helpful discussions.
AKM and AJB acknowledge support from NSF grants AST-1814421 and AST-2308161. AKM thanks the LSST-DA Data Science Fellowship Program, which is funded by LSST-DA, NSF Cybertraining Grant \#1829740, the Brinson Foundation, and the Moore Foundation; his participation in the program has benefited this work. AJB acknowledges support from the Radcliffe Institute for Advanced Study at Harvard University. EG acknowledges support from NSF grant AST-2206222. DO is a recipient of an Australian Research Council Future Fellowship (FT190100083) funded by the Australian Government. MG is supported by the Australian Government through the Australian Research Council's Discovery Projects funding scheme (DP210102103), by UK STFC Grant ST/Y001117/1, and by IDIA. MJJ acknowledges support from the Oxford Hintze Centre for Astrophysical Surveys  (funded through generous support from the Hintze Family Charitable Foundation), from Science and Technology Council (STFC) consolidated grants [ST/S000488/1] and [ST/W000903/1], and from a United Kingdom Research and Innovation (UKRI) Frontiers Research Grant [EP/X026639/1] selected by the European Research Council.
For the purpose of open access, the authors have applied a Creative Commons Attribution (CC BY) license to any Author Accepted Manuscript (AAM) version arising from this submission.

\newpage




\appendix

\section{Likelihood selection} \label{app:lkl}

In order to fit a model to the data or sample the posterior probability distribution of the model parameters using an MCMC method, we need to define a likelihood function. It is common practice to minimize the Mean Squared Error (MSE) between the data and the model, selecting the parameters that minimize the MSE as the best-fit model. Minimizing the MSE is equivalent to maximizing the Gaussian likelihood, under the assumption that the errors on the data are Gaussian-distributed and are determined based on the observed counts \citep{hogg10}. However, this assumption can lead to biased results when dealing with Poisson-distributed data, especially when counts are low. We summarize the inferred Schechter function parameters and their uncertainties for different likelihood functions in Table \ref{tab:lkl}. It is well-established that for large mean values, the Poisson distribution can be approximated as a Gaussian distribution with the same mean and variance; hence, for large mean values, maximizing the Gaussian likelihood is an excellent approximation to maximizing the Poisson likelihood. Similarly, the difference between minimizing the MSE and maximizing the Gaussian likelihood becomes small for large mean values. As shown in the likelihood column of Table \ref{tab:lkl}, the difference between the MSE and Gaussian likelihoods can be clarified by comparing how errors are calculated: Poisson error based on observed data\footnote{When errors are calculated based on observed data, the error term in the overall likelihood is a constant --- $\sum_{i = 1}^{N_{\rm bins}} \ln(y_i)$ using the terminology of Table \ref{tab:lkl} --- and can be factored out of the likelihood sum.} ($\sqrt{y}$ for an observed count of $y$) versus Poisson error based on model mean ($\sqrt{\mu}$ for a predicted mean of $\mu$). For instance, if the observed count deviates from the model mean by one standard deviation, i.e., $y = \mu + \sqrt{\mu}$, the relative error ($\frac{\sqrt{y}}{y}$) estimated from the observed data is smaller by a factor of $\sqrt{1+\frac{1}{\sqrt{\mu}}}$ compared to the relative error estimated from the mean ($\frac{\sqrt{\mu}}{\mu}$). This computation illustrates that for large mean values, the discrepancy between model-based and data-based error estimates becomes small (e.g., for $\mu=25$, the discrepancy for an observed count that deviates from the mean by one standard deviation is only $\sim 10\%$), making the MSE an effective approximation of the Gaussian likelihood. However, for cases with small means, the likelihood of larger-than-mean observations is significantly smaller for the Gaussian distribution compared to the Poisson distribution, and the full Poisson likelihood is required to ensure unbiased results. Using Gaussian or MSE likelihoods instead of a Poisson likelihood disproportionately influences the overall likelihood of the fit due to the impact of overfitting the bins with few detections; we therefore use the Poisson distribution in deriving our results with the recovery matrix method.

\begin{table*}
\centering
\caption{Inferred Schechter function parameters for different likelihood functions. $\mu_i$ and $y_i$ represent the predicted and observed counts, respectively.}
\label{tab:lkl}
\begin{tabular}{llccc}
\hline
 & Negative log likelihood  & $\phi_\ast \times 10^3$ (Mpc$^{-3}$) & log($M_\ast/M_\odot$) & $\alpha$    \\
\hline
MSE & $\sum_{i = 1}^{N_{\rm bins}} (\mu_i - y_i)^2/\mu_i $ & $2.87_{-1.17}^{+0.49}$ & $10.09_{-0.07}^{+0.23}$ & $-1.26_{-0.12}^{+0.05}$\\
Gaussian & $\sum_{i = 1}^{N_{\rm bins}} \ln(\mu_i) + (\mu_i - y_i)^2/\mu_i $ & $3.44_{-1.17}^{+0.66}$ & $9.99_{-0.08}^{+0.18}$ & $-1.21_{-0.11}^{+0.06}$ \\ 
Poisson & $\sum_{i = 1}^{N_{\rm bins}} \mu_i - \ln(\mu_i) y_i $ & $3.56_{-1.92}^{+0.97}$ & $10.01_{-0.12}^{+0.31}$ & $-1.18_{-0.19}^{+0.08}$ \\
\hline
\end{tabular}
\end{table*}

\section{Uncertainty due to cosmic variance} \label{app:unc}

To provide an indication of how cosmic variance affects our results, we use the ALFALFA $\alpha.100$ \hi\ source catalog \citep{haynes18} to estimate how many high-mass \hi\ galaxies would be detected within volumes comparable to that of LADUMA's low-$z$ SPW cube. We begin by choosing a subsample of the ALFALFA catalog with high completeness, by excluding sources with log(\mhi/$M_\odot$)$< 9.75$ and $z>0.05$. These cuts minimize the impact of survey incompleteness and result in a sample of over 8500 galaxies. To ensure a fair comparison with the low-$z$ SPW for LADUMA, we define 388 independent ``pencil-beam'' sub-volumes within the ALFALFA volume; each of these is matched to the volume of the LADUMA low-$z$ SPW by selecting a larger solid angle on the sky to compensate for the smaller redshift coverage of ALFALFA. Within each defined sub-volume, we count the number of galaxies. We then calculate the standard deviation in galaxy counts across these sub-volumes, which is found to be $\sim 8.5$ (relative to a mean of 16). This observed standard deviation exceeds the expected Poisson uncertainty of $\sqrt{16} = 4$, providing an indication of the impact of cosmic variance on our results. In our LADUMA sample, the observed number of high-mass sources is 15, closely matching the average number derived from the ALFALFA test.  While this exercise suggests that Poisson uncertainties represent only $\sim 47\%$ of the total uncertainty when cosmic variance is included, the facts that uncertainties are not symmetrically distributed for the Schechter function parameters and that the ALFALFA sample does not provide a sufficiently robust estimate of cosmic variance for low-mass galaxies (due to the smaller available volumes) mean that extending this analysis to individual Schechter function parameters is non-trivial.

\section{Comparing maximum volume and injection/recovery methods} \label{app:veff}

Considering the intrinsic number density, $\phi_t$, of sources with \hi mass $M$, i.e., the value of the \himf at $M_{\rm HI} = M$,  we can write:
\begin{equation} \label{eq:tr_phi}
    \phi_t(M)\,d{\rm log}\,M = \frac{n_t}{V_s},
\end{equation}
where $d{\rm log}\,M$ is the width of a bin centered at $M$ and $n_t$ is the intrinsic number of sources (not necessarily an integer) within a survey's volume, denoted by $V_s$\footnote{We use a $t$ subscript on $\phi_t$ and $n_t$ to evoke the ``true" values of the mass function and the number of sources.}. We can express the observed number density of sources with \hi mass $M$ in a survey using a $V_{\rm max}$ (VM) framework as
\begin{equation} \label{eq:vm_phi}
    \phi_{\rm VM}(M)\,d{\rm log}\,M = \frac{n_o}{V_{\rm eff}}
\end{equation}
and using an injection/recovery (IR) framework as
\begin{equation} \label{eq:ir_phi}
    \phi_{\rm IR}(M)\,d{\rm log}\,M = \frac{n_o}{C_s V_s},
\end{equation}
where $n_o$ represents the integer number of sources detected in an interval of width $d{\rm log}\,M$ centered at $M$, which can be described as a randomly sampled (integer) value from a Poisson distribution with mean $\mu = C_s n_t$, i.e., $n_o \sim P(\mu = C_s n_t)$; $V_{\rm eff}$ is the effective volume of sources with \hi mass $M$ (estimated using the observed sample); and $C_s$ is the overall completeness of the survey for all types of sources with \hi mass $M$ (estimated via injection/recovery tests). Comparing Equations \ref{eq:vm_phi} and  \ref{eq:ir_phi} shows that the $V_{\rm max}$ approach estimates the completeness of a survey by calculating the effective volume ($V_{\rm eff}$) using the observed source population. It is clear that any error introduced by the randomness of the observed population affects the numerators in Equations \ref{eq:vm_phi} and \ref{eq:ir_phi} identically. For the denominators, this analogy breaks down. The denominator in Equation \ref{eq:ir_phi} is not affected by errors due to the randomness of the observed population, as $V_s$ is fixed and the completeness $C_s$, is determined via injection/recovery tests. However, the denominator in Equation \ref{eq:vm_phi} --- when determined using the observed population --- will be affected by errors due to the randomness in the observed sample. It is helpful to track how random Poisson uncertainties affect the observed number density for these two methods. For this exercise, we assume that the uncertainties introduced by the injection/recovery process are negligible (as they can be reduced by increasing the number of injected sources) compared to the random uncertainties.

First, we need to consider the different ``types" of galaxies with $M_{\rm HI} = M$. We assume there are $N$ types of galaxies with normalized abundances of $x_i$, where $\Sigma_{i=1}^N x_i = 1$; one example would be galaxies with the same mass but different inclination angles. Assuming a completeness fraction of $C_i$\footnote{This completeness fraction $C_i$ for sources of type $i$ can be interpreted as the fraction of the total volume available for the detection of sources of type $i$, i.e., $V_i = C_i V_s$.} for each source type, the expected number of sources $\langle n_o \rangle$ in the survey volume would be
\begin{align}
    \langle n_o \rangle &= \Sigma_{i=1}^N \phi_t x_i C_i V_s \\
    &= \Sigma_{i=1}^N \phi_i V_i = \Sigma_{i=1}^N \langle n_i \rangle \label{eq:ln2}\\
    &= \phi_t V_s \Sigma_{i=1}^N x_i C_i = \phi_t V_s C_s = n_t C_s, \label{eq:ln3}
\end{align} 
where $\phi_i = \phi_t x_i$, $V_i = C_i V_s$, and $C_s = \Sigma_{i=1}^N x_i C_i$ are respectively the intrinsic number density of sources of type $i$, the effective volume available for the detection of sources of type $i$, and the completeness of the survey for sources of \hi mass $M$, and $\langle n_i \rangle$ is the expected number of sources of type $i$ in the survey volume. Equations \ref{eq:ln2} and \ref{eq:ln3} are two representations of the expected number of sources: the sum of the expected numbers of sources of type $i$ (\ref{eq:ln2}) and the intrinsic number of sources multiplied by the completeness of the survey (\ref{eq:ln3}). The number of observed sources $n_o$ can be expanded as
\begin{equation} \label{eq:expand}
    n_o = \langle n_o \rangle + \delta n_o = \Sigma_{i=1}^N \langle n_i \rangle + \Sigma_{i=1}^N \delta n_i = \Sigma_{i=1}^N (\langle n_i \rangle + \delta n_i) = \Sigma_{i=1}^N n_i, 
\end{equation}
where $\delta n_o$ and $\delta n_i$ (which may be positive or negative) are respectively the random uncertainty in the total number of observed sources and the random uncertainty in the number of observed sources of type $i$, and $n_i$ is the number of observed sources of type $i$\footnote{The sum, $Z = X+Y $, of two independent Poisson random variables, $X \sim {\rm Poisson}(\mu_x)$ and $Y \sim {\rm Poisson}(\mu_y)$, is also a Poisson random variable, $Z \sim {\rm Poisson}(\mu_x+\mu_y)$.}. 

In an injection/recovery framework, by combining equations \ref{eq:ir_phi}, \ref{eq:ln3}, and \ref{eq:expand}, we have:
\begin{align} 
    \phi_{\rm IR}(M)\,d{\rm log}\,M &= \frac{n_o}{C_s V_s} = \frac{\langle n_o \rangle + \delta n_o}{C_s V_s} = \frac{\langle n_o \rangle}{C_s V_s} + \frac{\delta n_o}{C_s V_s} = \frac{\phi_t C_s V_s}{C_s V_s} + \frac{\delta n_o}{C_s V_s} \label{eq:ir_def}\\
    & = \phi_t + \frac{\delta n_o}{C_s V_s} \label{eq:ir_last}, 
\end{align}
showing that the uncertainty in the observed number density is proportional to the random uncertainty in the total number of observed sources with $M_{\rm HI} = M$. When the number of observed sources is equal to the expected number of sources $n_o = \langle n_o \rangle$, i.e., $\delta n_o = \Sigma_{i=1}^N \delta n_i  = 0$, the observed number density is equal to the intrinsic number density. Similarly, in a $V_{\rm max}$ framework, we can combine equations \ref{eq:vm_phi}, \ref{eq:ln2}, and \ref{eq:expand} to get:
\begin{align}
    \phi_{\rm VM}(M)\,d{\rm log}\,M &= \frac{n_o}{V_{\rm eff}} = \Sigma_{i=1}^N \frac{n_i}{V_i} = \Sigma_{i=1}^N \frac{\langle n_i \rangle + \delta n_i}{V_i} \label{eq:vm_def}\\
    & = \Sigma_{i=1}^N \frac{\langle n_i \rangle}{V_i}  + \Sigma_{i=1}^N \frac{\delta n_i}{V_i} =  \Sigma_{i=1}^N \frac{\phi_i V_i}{V_i}  + \Sigma_{i=1}^N \frac{\delta n_i}{V_i} \\
    & =  \Sigma_{i=1}^N x_i \phi_t  + \Sigma_{i=1}^N \frac{\delta n_i}{V_i} =  \phi_t \Sigma_{i=1}^N x_i  + \Sigma_{i=1}^N \frac{\delta n_i}{V_i} \\
    &=  \phi_t + \Sigma_{i=1}^N \frac{\delta n_i}{V_i} \label{eq:last},
\end{align}
which shows that the uncertainty in the measured number density is proportional to the sum of the random uncertainties in the numbers of observed sources with $M_{\rm HI} = M$ of type $i$ divided by the effective volumes for sources of type $i$. Equation \ref{eq:last} shows that when $\Sigma_{i=1}^N \frac{\delta n_i}{V_i} = 0$, the measured number density would be equal to the intrinsic number density. The trivial case is when $\delta n_i = 0$ for all source types, i.e., for all source types, the number of observed sources of type $i$ is equal to expected number of sources of type $i$,  $n_i = \langle n_i \rangle $. The more complicated scenario will be for all the nonzero $\{\delta n_i\}$ in the sum to cancel each other. An interesting case is to compare equations \ref{eq:ir_last} and \ref{eq:last} in the limiting case of a volume-limited sample for all sources of type $i$. For a volume-limited sample, the effective volume for each source type is equal to the total volume of the survey, i.e., $V_i = V_s$, and the completeness for sources of type $i$ is equal to 1, as all such sources can be detected throughout the total volume, i.e., $C_i = 1$. Substitution of these values into equations \ref{eq:ir_last} and \ref{eq:last} results in
\begin{align}
    \ref{eq:ir_last} \rightarrow~& \phi_{\rm IR}(M)\,d{\rm log}\,M = \phi_t + \frac{\delta n_o}{C_s V_s} = \phi_t + \frac{\delta n_o}{V_s}  \\
    \ref{eq:last} \rightarrow~& \phi_{\rm VM}(M)\,d{\rm log}\,M =  \phi_t + \Sigma_{i=1}^N \frac{\delta n_i}{V_i}=  \phi_t + \Sigma_{i=1}^N \frac{\delta n_i}{V_s} =  \phi_t +  \frac{ \Sigma_{i=1}^N \delta n_i}{V_s} = \phi_t + \frac{\delta n_o}{V_s} ,
\end{align}
making it clear that when all types of sources of \hi mass $M$ are detectable throughout the entire volume of the survey, the two approaches are equivalent. We note that in these derivations we have assumed that the effective volume and the completeness of the survey are perfectly determined, underscoring the fact that the difference between these two approaches arise from the methods themselves and are not due to imperfect calculations. In practice, the effective volume and the completeness fraction calculations are affected by errors that contribute to the overall uncertainties in the measured number density distributions.

\begin{figure}
\centering
\includegraphics[width=0.55\textwidth]{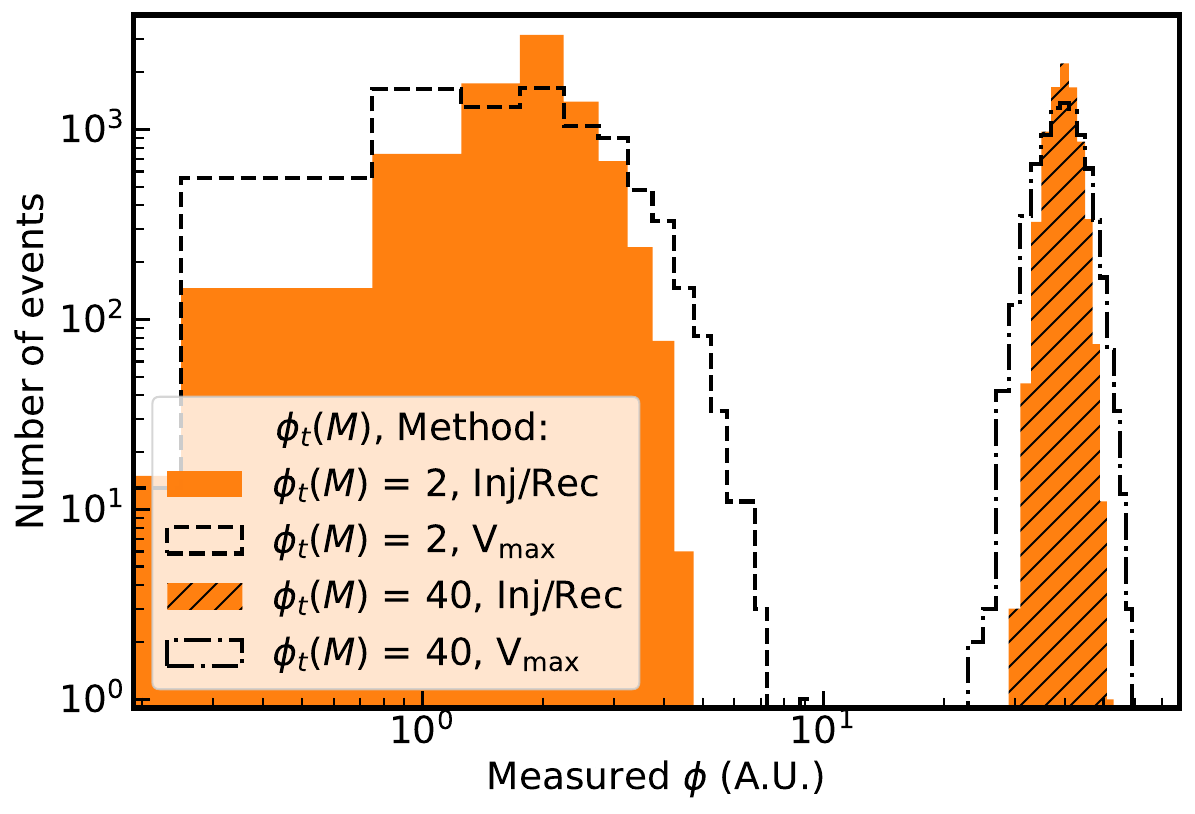}
\caption{Distributions of observed number density of sources with $M_{\rm HI} = M$ for the $V_{\rm max}$ and injection/recovery methods. It is clear that the $V_{\rm max}$ method leads to a wider distribution of observed number densities around the true number density of sources with $M_{\rm HI} = M$. This effect is due to the fact that the $V_{\rm max}$ method uses the observed dataset to estimate the completeness of the survey and therefore is affected by uncertainties in the observed population to a larger degree than the injection/recovery method. It is important to note that these differences are inherent in the methods themselves, and this figure assumes perfect knowledge of the completeness of the survey and the effective volume for the (two) different types of galaxies of \hi mass $M$.
}
\label{fig:methods}
\end{figure}

Considering a simplified case of two types of galaxies of equal abundance and identical \hi mass $M$, following an approach similar to Appendix C of \citet{obreschkow18}, we demonstrate how random uncertainties affect the observed number density of sources with $M_{\rm HI} = M$. Assuming that $\phi_t(M) = 2$, $x_1=x_2=0.5$, $C_1 = 1/8$, $C_2 = 1$, and $V_s = 8$, we can calculate the expected number of galaxies as
\begin{align} \label{eq:expected}
    \langle n_o \rangle &= \Sigma_{i=1}^N \langle n_i \rangle\\
    &= \Sigma_{i=1}^N \phi_t x_i C_i V_s \\
    & = \phi_t x_1 C_1 V_s +  \phi_t x_2 C_2 V_s = \langle n_1 \rangle + \langle n_2 \rangle \\
    &= 2\cdot \frac{1}{2}\cdot \frac{1}{8} \cdot 8 + 2\cdot \frac{1}{2}\cdot 1 \cdot 8 \\
    & = 1 + 8 = 9.
\end{align}
Therefore, we expect to observe one galaxy of type 1 and eight galaxies of type 2, totaling to nine galaxies of $M_{\rm HI} = M$ in the survey volume. We can randomly sample two Poisson distributions with means of $\mu_1 = 1$ and $\mu_2 = 8$ to simulate observations and calculate the observed number density of sources with $M_{\rm HI} = M$ for the two approaches using equations \ref{eq:vm_def} and \ref{eq:ir_def}. A histogram of the outcome of $2^{13}$ simulated observations for cases of $\phi_t(M) = 2$ and $\phi_t(M) = 40$ is shown in Figure \ref{fig:methods}, demonstrating that the $V_{\rm max}$ approach leads to a wider distribution of observed number densities around the true number density of sources with $M_{\rm HI} = M$, due to the fact that the $V_{\rm max}$ approach uses the observed dataset to estimate the completeness of the survey and therefore is affected by the uncertainties in the observed population to a larger degree compared to the injection/recovery approach. 

We close by noting that the preceding analysis treats the parameter ($M$) whose distribution (the \himf) we are trying to derive in a fundamentally different way from other parameters (e.g., inclination) that determine what ``types'' different galaxies have, in the sense of affecting the maximum volumes in which they can be detected. This distinction is not absolute: if we were interested (for example) in deriving the \hi velocity width function, we could do so treating $M$ as a parameter that determines ``type,'' as it affects the effective available volume for the detection of sources of a given velocity width, and the limitations of the $V_{\rm max}$ approach relative to the injection/recovery approach would remain the same as we have characterized them above.


\bibliography{laduma_himf}{}
\bibliographystyle{aasjournal}



\end{document}